\documentclass[showpacs,preprintnumbers,nofootinbib,
superscriptaddress,amsmath,floatfix,prd]{revtex4}

\usepackage{amssymb}  
\usepackage{graphicx}

\usepackage[dvips]{color}

\usepackage[normalem]{ulem}  

\newcommand{\non}{\nonumber\\}

\newcommand{\be}{\begin{equation}}
\newcommand{\ee}{\end{equation}}
\newcommand{\bea}{\begin{eqnarray}}
\newcommand{\eea}{\end{eqnarray}}
\newcommand{\ba}[1]{\begin{array}{#1}}
\newcommand{\ea}{\end{array}}

\begin{document}

\title{Bulk viscosity in kaon-condensed color-flavor locked quark matter}

\author{Mark G.\ Alford}
\email{alford@wuphys.wustl.edu}
\affiliation{Department of Physics, Washington University St Louis, MO, 63130, USA}

\author{Matt Braby}
\email{mbraby@hbar.wustl.edu}
\affiliation{Department of Physics, Washington University St Louis, MO, 63130, USA}

\author{Andreas Schmitt}
\email{aschmitt@hep.itp.tuwien.ac.at}
\affiliation{Department of Physics, Washington University St Louis, MO, 63130, USA}
\affiliation{Institut f\"{u}r Theoretische Physik, Technische Universit\"{a}t Wien, 1040 Vienna, Austria}

\date{September 3, 2008}

\begin{abstract}

Color-flavor locked (CFL) quark matter at high densities is a color
superconductor, which spontaneously breaks baryon number and chiral
symmetry.  Its low-energy thermodynamic and transport properties are
therefore dominated by the $H$ (superfluid) boson, and the octet of
pseudoscalar pseudo-Goldstone bosons of which the neutral kaon is the
lightest. We study the CFL-$K^0$ phase, in which the stress induced by
the strange quark mass causes the kaons to condense, and there is an
additional ultra-light ``$K^0$'' Goldstone boson arising from the
spontaneous breaking of isospin.  We compute the bulk viscosity of
matter in the CFL-$K^0$ phase, which arises from the
beta-equilibration processes $K^0 \leftrightarrow H+H$ and $K^0 +H
\leftrightarrow H$. We find that the bulk viscosity varies as $T^7$,
unlike the CFL phase where it is exponentially Boltzmann-suppressed by
the kaon's energy gap. However, in the temperature range of relevance
for $r$-mode damping in compact stars, the bulk viscosity in the
CFL-$K^0$ phase turns out to be even smaller than in the uncondensed
CFL phase, which already has a bulk viscosity much smaller than all
other known color-superconducting quark phases.

\end{abstract}


\maketitle

\section{Introduction}
\label{intro}

The color-flavor locked (CFL) phase of quark matter is the densest predicted state of matter \cite{Alford:2007xm};
it may occur in nature, in the core of compact stars, which are expected to reach
several times nuclear saturation density. Quark matter is described by the theory of 
Quantum Chromodynamics (QCD), which, because of  asymptotic freedom \cite{Gross:1973id,Politzer:1973fx}, 
becomes weakly coupled and hence perturbatively tractable at asymptotically high densities. In that
regime, the CFL phase can be shown to be the ground state. However, at compact-star densities
the coupling is strong, so first-principles calculations are not possible.
We therefore follow a more phenomenological approach. We start with
the hypothesis that color-flavor-locked quark matter occurs in compact stars, and
calculate phenomenologically relevant properties which could allow us to find
astrophysical signatures of its presence. We will calculate the bulk viscosity,
which is relevant for the damping of pulsations and (indirectly) the spin down
of the star.

If the interior of the star is a perfect (dissipationless) fluid, then a certain class of oscillating modes, $r$-modes, 
are unstable with respect to the emission of gravitational waves \cite{Andersson:1997xt,Lindblom:2000jw,Owen:1998xg}. This 
emission acts as a brake on the rotation of the star. Since we know from observations
that there are compact stars with very large rotation frequencies, we conclude that the instability must be damped. One
damping mechanism is a nonzero viscosity of the fluid. Both shear and bulk viscosity can affect the $r$-modes and typically
act in different temperature regimes. Therefore, it is of physical interest to compute shear and bulk viscosity of various
candidate phases in a compact star as a function of temperature. Several calculations exist in the literature, for nuclear 
\cite{Haensel:2000vz,Haensel:2001mw,Andersson:2004aa,Chatterjee:2007qs,Gusakov:2007px,Chatterjee:2007iw,Chatterjee:2007ka} and hyperonic 
\cite{Lindblom:2001hd,Haensel:2001em} as well as for unpaired quark matter \cite{Madsen:1999ci,Madsen:1992sx,Sa'd:2007ud} and various 
color-superconducting phases \cite{Manuel:2004iv,Alford:2006gy,Sa'd:2006qv,Manuel:2007pz}, for a review see \cite{Dong:2007ax}. Bulk viscosity
due to thermal kaons in the CFL phase has been computed in Ref.\ \cite{Alford:2007rw}. Here, we extend this 
calculation to the case of condensed kaons.

At asymptotically large quark chemical potentials $\mu_q$, three-flavor quark matter is in the CFL phase \cite{Alford:1998mk}. 
This phase breaks the local color gauge group $SU(3)_c$, giving rise to Meissner masses for all eight gluons. It also breaks the global chiral 
symmetry group $SU(3)_L\times SU(3)_R$, leaving a residual global $SU(3)_{c+L+R}$ which contains simultaneous rotations in color and flavor 
space, hence ``locking'' color with flavor. Moreover, the global baryon number conservation symmetry $U(1)_B$ is broken, rendering 
the CFL phase a superfluid. For massless quarks there are thus 8+1 massless Goldstone bosons (and an additional Goldstone boson for the $U(1)_A$
symmetry, which, however, is expected to be explicitly broken at moderate chemical potentials). At moderate densities the mass of the 
strange quark $m_s$ cannot be neglected and the bosons of the octet associated with chiral symmetry breaking acquire masses. These masses
are small compared to the energy gaps of the fermions, and therefore the low-energy properties of the CFL phase can be described
within an effective theory for the Goldstone bosons \cite{Son:1999cm,BedaqueSchaefer,Kaplan:2001qk}. 
We shall make use of this theory in this paper.
If $m_s^2/\mu_q$ is large enough,
it is expected that the lightest pseudo-Goldstone bosons, namely the neutral kaons, condense. The CFL phase with $K^0$ condensation
is called the ``CFL-$K^0$'' phase.
Goldstone bosons in CFL and their condensation have also been studied in a different approach, using a Nambu-Jona-Lasinio model 
\cite{Forbes:2004ww,Buballa:2004sx,Warringa:2006dk,Ruggieri:2007pi,Ebert:2007bp,Kleinhaus:2007ve}.

Besides giving rise to masses for the meson octet, a nonzero strange mass induces a mismatch in the Fermi momenta
of the quarks that form Cooper pairs in the CFL phase. In fact, the strange mass induces a mismatch in any possible spin-zero
color-superconducting phase 
\cite{Rajagopal:2005dg}. This means that at lower densities, the particularly symmetric CFL phase may be replaced by a less symmetric 
pairing pattern. It is currently not known whether, going down in density, the CFL phase is superseded by nuclear matter or by a different, more
exotic, color-superconducting phase. Candidate color-superconducting phases have Cooper pairs with nonzero angular momentum 
\cite{Schafer:2000tw,Alford:2002rz,Schmitt:2002sc,Schmitt:2004et} or nonzero momentum 
\cite{Alford:2000ze,Rajagopal:2006dp,Mannarelli:2006fy}. In this paper, we shall only consider the CFL 
and CFL-$K^0$ phases.

The paper is organized as follows. In Sec.\ \ref{low} we give a brief overview over the properties of the CFL-$K^0$ phase, in particular
we present the low-energy excitations at finite temperature. As an application, we discuss the resulting specific heat of the system in 
Sec.\ \ref{specific}. The calculation of the bulk viscosity is presented in Sec.\ \ref{bulk}. We define the bulk viscosity 
in Sec.\ \ref{definition}. In Secs.\ \ref{density} and \ref{rates} we collect the ingredients needed for the bulk viscosity, namely the
kaon density and susceptibility and the rate of the processes $K^0 \leftrightarrow H+H$ and $K^0 +H \leftrightarrow H$. We put these 
ingredients together in Sec.\ \ref{results} to obtain the result for the bulk viscosity and give our conclusions in Sec.\ \ref{conclusions}.  

\section{Low-energy modes in the CFL-$K^0$ phase}
\label{low}

In this section, we briefly summarize the theoretical description and physical properties of the Goldstone bosons in the CFL-$K^0$ phase. 
More details can be found in the references given in the text.

\subsection{Chiral Lagrangian} 
\label{chiral}

We denote the meson nonet in the CFL phase, associated with chiral symmetry breaking, by 
\be
\Sigma = e^{i\theta/f_\pi} \, ,  
\ee
where $\theta$ is an element of the Lie algebra of $U(3)$. 
The physics of the mesons is described by the Lagrangian \cite{BedaqueSchaefer,Son:1999cm}
\bea \label{Leff}
{\cal L} &=& \frac{f_\pi^2}{4} {\rm Tr}\left[(\partial_0\Sigma+i[A_0,\Sigma])
(\partial_0\Sigma^\dag-i[A_0,\Sigma]^\dag) - v_\pi^2\partial_i\Sigma\partial_i\Sigma^\dag\right]
+\frac{af_\pi^2}{2}{\rm det}\hat{M}{\rm Tr}[\hat{M}^{-1}(\Sigma+\Sigma^\dag)] \, .
\eea
There is an effective chemical potential given by the ``gauge field''
\be
A_0\equiv \mu_Q Q -\frac{\hat{M}^2}{2\mu_q} \, ,
\ee
where $Q={\rm diag}(2/3,-1/3,-1/3)$ and $\hat{M}={\rm diag}(m_u,m_d,m_s)$ are the electric charge and quark 
mass matrices in flavor space, and $\mu_Q$ is the chemical potential associated with electric charge.
Moreover, from matching calculations at asymptotically large densities we know
\bea \label{matching}
f_\pi^2 &=& \frac{21-8\ln 2}{18}\frac{\mu_q^2}{\pi^2} \, , \qquad v_\pi = \frac{1}{\sqrt{3}} \, , \qquad
a=\frac{3\Delta^2}{\pi^2f_\pi^2} \, ,
\eea
where $\Delta$ is the fermionic energy gap at zero temperature. It turns out that the neutral and charged kaons are the lightest
mesons. They carry flavor quantum numbers $K^0\sim \bar{s}d$, $\bar{K}^0\sim \bar{d}s$, 
$K^+\sim \bar{s}u$, $K^-\sim \bar{u}s$. In contrast
to the usual mesons, however, they are composed of four quarks of the structure $\bar{q}\bar{q}qq$ as opposed to $\bar{q}q$. 
The zero-temperature kaon masses and effective chemical potentials are deduced from the Lagrangian and are given by   
\begin{subequations} \label{masschem}
\bea
\mu_{K^+}&\equiv&\mu_Q+\frac{m_s^2-m_u^2}{2\mu} \, , \qquad \mu_{K^0} \equiv \frac{m_s^2-m_d^2}{2\mu} \, , \\
m_{K^+}^2&\equiv& am_d(m_s+m_u) \, , \qquad m_{K^0}^2 \equiv  am_u(m_s+m_d)
\, .
\eea
\end{subequations}
We see that with $m_d$ slightly larger than $m_u$ the neutral kaon is slightly lighter than the charged kaon. Moreover, electric
neutrality disfavors the presence of charged kaons. Therefore, in most of the remainder of the paper we shall ignore the charged kaons.
In the following, for notational convenience, we denote the neutral kaon chemical potential and mass simply by $\mu$ and $m$, respectively,
\be
\mu\equiv \mu_{K^0} \, , \qquad m\equiv m_{K^0} \, .
\ee 
Condensation of 
the neutral kaons occurs if $\mu>m$. Because of the large uncertainty in the quark masses and in the dimensionless quantity 
$a$, it is not clear whether
this condition is fulfilled for densities present in the interior of a compact star. Using the high-density expressions 
in Eq.\ (\ref{matching}) and inserting a quark chemical potential $\mu_q\simeq 500\,{\rm MeV}$ and an energy gap $\Delta\simeq 30\,{\rm MeV}$, we estimate 
$f_\pi\simeq 100\,{\rm MeV}$, $a\simeq 0.01$. With $m_d\simeq 7\,{\rm MeV}$, $m_u\simeq 4\,{\rm MeV}$ we thus obtain a kaon chemical potential $\mu\simeq 20\,{\rm MeV}$ 
and a kaon mass $m\simeq 4\,{\rm MeV}$. These values suggest that the kaons are condensed and thus that the relevant phase to consider 
is the CFL-$K^0$ phase. It has been argued that for sufficiently large values of the parameter $m_s^2/\mu_q$ the CFL-$K^0$ phase is modified
to the so-called curCFL-$K^0$ phase which is anisotropic and exhibits counter-propagating currents from the kaon condensate and ungapped fermions 
\cite{Kryjevski:2008zz,Schafer:2005ym,Gerhold:2006np}. Because of the presence of ungapped fermions, the transport properties 
of this phase can be expected to be very different from the ones in the CFL-$K^0$ phase and similar to unpaired quark matter. 
In this paper, we do not consider the possibility of a kaon current but rather focus on the isotropic CFL-$K^0$ phase where all 
fermions are gapped. 

As we shall see in Sec.\ \ref{bulk}, in order to compute the bulk viscosity we need the nonzero-temperature behavior of the kaons, 
in particular their masses and excitation energies. The necessary results are provided in Ref.\ \cite{Alford:2007qa} and shall 
be briefly summarized in the next two subsections. 

\subsection{2PI formalism for kaons}

To obtain the Lagrangian for the neutral kaons in the presence of a kaon condensate we start from the Lagrangian (\ref{Leff}) and set 
\be \label{sigma67}
\Sigma=e^{i[(\phi+\varphi_6)T_6+\varphi_7T_7]/f_\pi} \, , 
\ee
where $T_6$, $T_7$ are Gell-Mann matrices. We have omitted the other Goldstone modes, proportional to ${\bf 1},T_1,\ldots,T_5,T_8$, 
and have introduced a vacuum expectation value for the kaon field $\phi$ (without loss of generality in the $T_6$ direction) and 
fluctuations $\varphi_6$, $\varphi_7$ with vanishing expectation value. The Lagrangian is now expanded up to fourth order 
in the fluctuations. See Ref.\ \cite{Alford:2007qa} for details and the explicit form of the Lagrangian. The
tree-level potential is 
\be \label{Uphi}
U(\phi) = f_\pi^2\left[m^2\left(1-\cos\frac{\phi}{f_\pi}\right)-\frac{\mu^2}{2}\sin^2\frac{\phi}{f_\pi}\right]
\simeq \frac{m^2-\mu^2}{2}\phi^2+\frac{\alpha}{4}\phi^4 \, ,
\ee
where we have abbreviated the effective coupling constant
\be \label{alpha}
\alpha\equiv\frac{4\mu^2-m^2}{6f_\pi^2} \, .
\ee
In principle, one can keep all orders in $\phi$, but for simplicity we have expanded for small values of $\phi$. This restricts
our analysis to small condensates. In other words, our results are quantitatively reliable if $\mu$ is only slightly larger 
than $m$ and become unreliable for $\mu\gg m$. In addition to the tree-level potential, the Lagrangian contains 
quadratic terms in the fluctuations from which we read off the inverse tree-level propagator
\bea \label{S0}
S_{0}^{-1}(K) &=& \left(\begin{array}{cc} -k_0^2+v_\pi^2k^2 + m^2-\mu^2+3\alpha\phi^2  & 
-2i\mu k_0 \\[1ex] 2i\mu k_0 &
-k_0^2+v_\pi^2k^2 + m^2-\mu^2+\alpha\phi^2  \end{array}\right) \, .
\eea
Here and in the following we denote four-momenta by capital letters $K=(k_0,{\bf k})$, where the bosonic Matsubara frequencies are given by 
$k_0=-2in\pi T$ with the temperature $T$. The Lagrangian also contains interaction terms cubic and quartic in the fluctuations. 
For small condensates we may neglect the cubic interactions. The form 
of the quartic interactions is needed for the vertex in the kaon self-energy.

\begin{figure} [t]
\begin{center}
\includegraphics[width=0.7\textwidth]{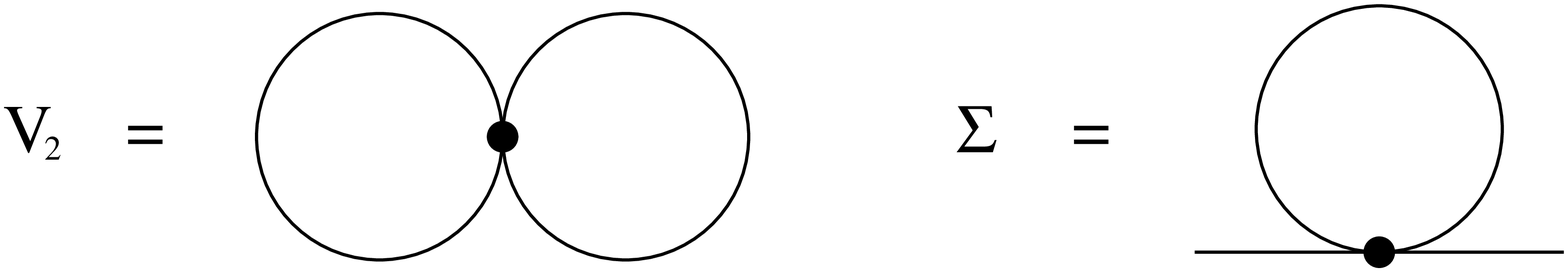}
\caption{
Diagrammatic representation of the two-loop approximation for $V_2$ and corresponding kaon self-energy. 
The lines represent the full kaon propagator $S$, to be determined self-consistently.
}
\label{figv2}
\end{center}
\end{figure}

The nonzero-temperature behavior of systems with spontaneously broken symmetries can be conveniently treated within the
two-particle irreducible (2PI) formalism \cite{Luttinger:1960ua,Baym:1962sx,Cornwall:1974vz}; for the application of this
formalism in chiral models see for instance Refs.\ \cite{Lenaghan:1999si,Roder:2003uz,Andersen:2006ys}. Naive thermal corrections 
would lead to unphysical results for a Bose-condensed system, in particular to negative energies for certain temperature regimes. Therefore,
we apply the more elaborate  2PI scheme to compute the thermal masses self-consistently. We start from the effective potential  
\be \label{Veff}
V_{\rm eff}[\phi,S] = U(\phi) + \frac{1}{2}{\rm Tr}\ln S^{-1} + 
\frac{1}{2}{\rm Tr}[S_0^{-1}(\phi)S-1] + V_2[\phi,S] \, ,
\ee
where the trace is taken over momentum space and over the two-dimensional space given by the two degrees of freedom of the kaon (corresponding 
to $T_6$ and $T_7$ or $K^0$ and $\bar{K}^0$).   
The effective potential is a functional of the vacuum expectation value $\phi$ and the kaon propagator $S$. 
The term $V_2[\phi,S]$ is the sum over all 2PI
diagrams. We employ the two-loop approximation for this infinite sum and only consider the ``double-bubble'' diagram, see Fig.\ \ref{figv2}. 
In this case $V_2$ does not explicitly depend on $\phi$. 

One can now determine the condensate and the thermal kaon mass self-consistently. This is done by solving
the stationarity equations 
 \begin{subequations}
\bea
0 &=& \frac{\partial U}{\partial \phi} + \frac{1}{2}{\rm Tr}\left[\frac{\partial S_0^{-1}}{\partial\phi} S
\right] \, , \label{condmini}\\   
S^{-1} &=& S_0^{-1} + \Sigma \, ,\label{schwinger}
\eea
\end{subequations}
where $\Sigma=2\frac{\delta V_2}{\delta S}$ is the kaon self-energy, see Fig.\ \ref{figv2}. With the ansatz for the full inverse propagator
\bea \label{fullprop}
S^{-1} &=& \left(\begin{array}{cc} -k_0^2+v_\pi^2k^2 + M_{+}^2-\mu^2  & 
-2i\mu k_0 \\[1ex] 2i\mu k_0 &
-k_0^2+v_\pi^2k^2 + M_{-}^2-\mu^2  \end{array}\right) \, , 
\eea
the stationarity equations read
\begin{subequations} \label{stationarity}
\bea
(M^2-\mu^2)\phi&=&\alpha\phi^3\, , \label{stat1}\\
M^2&=&m^2+2\alpha\phi^2+2\alpha I(M^2,T) \, . \label{stat2}
\eea
\end{subequations}
Here, $M$ is the temperature dependent mass, related to $M_\pm$ in the propagator (\ref{fullprop}) by 
$M^2=(M_+^2+M_-^2)/2$. The other combination $\delta M^2\equiv (M_+^2-M_-^2)/2$ can be eliminated since the second matrix component 
of Eq.\ (\ref{schwinger}) implies $\delta M^2 \simeq \alpha \phi^2$.\footnote{In fact we have $\delta M^2 = 
\alpha \phi^2/(1+\alpha\,J)$ with $J$ being a momentum integral similar to $I$ in Eq.\ (\ref{defI}). We may neglect the term $\alpha\,J$. 
This removes a very small mass for the Goldstone boson, which
arises as an artefact in the 2PI formalism at finite
temperature. For details and the expression for $J$ see Ref.\ \cite{Alford:2007qa}.} 
We have abbreviated
\be \label{defI}
I(M^2,T)\equiv \int\frac{d^3{\bf k}}{(2\pi)^3}\frac{f(\epsilon_k)}{E_k} \, ,
\ee
with 
\be \label{defEk}
E_k\equiv \sqrt{v_\pi^2k^2+M^2} \, ,
\ee
and the Bose distribution function 
\be \label{bose}
f(x)\equiv \frac{1}{e^{x/T}-1} \, .
\ee
The kaon excitation energy $\epsilon_k$ is a pole of the propagator (\ref{fullprop}). Its form is crucial for the thermodynamic and  
transport properties of the CFL-$K^0$ phase and will be discussed explicitly in Sec.\ \ref{kaonenergy}, see Eq.\ (\ref{excite}). 
The second pole of the propagator corresponds to the $\bar{K}^0$ excitation. This excitation has a (temperature-dependent) energy gap 
which is larger than the $K^0$ excitation by at least $2\mu$. We will neglect the $\bar{K}^0$ mode in our analysis, because
$\mu$ is expected to be on the order of $\sim 10\,{\rm MeV}$, so the $\bar{K}^0$ would only play a role at temperatures in the
tens of MeV range. This is close to the critical temperature of the CFL phase itself, and is not physically
relevant because compact stars cool below such temperatures within minutes of their formation. Our results are therefore reliable
at temperatures of order 1~MeV or lower. 
In some cases we will continue our expressions to higher temperatures: this represents
a theoretical exercise in studying the properties of the pure
$K^0$ condensate, not a physical prediction.

For the calculations of the specific heat in Sec.\ \ref{specific} as well as for the density and susceptibility in Sec.\ \ref{density} we
need an explicit expression for the pressure $P$ which is the   
negative of the effective potential (\ref{Veff}), $P=-V_{\rm eff}$. In the given approximation we can write $P$ as a function of $\phi^2$ and $M^2$,
\be \label{pressure}
P(\phi^2,M^2)=-U(\phi^2)-T\int\frac{d^3{\bf k}}{(2\pi)^3}\ln\left(1-e^{-\epsilon_k/T}\right)
+\frac{1}{2}(M^2-m^2-2\alpha\phi^2)\,I(M^2,T) -\frac{\alpha}{2} I^2(M^2,T)\, .
\ee
The pressure in the physical state is then given by inserting the values for $\phi$ and $M$ at the stationary point.

\subsection{Kaon excitation energies}   
\label{kaonenergy}

For the calculation of transport properties we are interested in the kaon dispersion relations which are the poles of the 
propagator (\ref{fullprop}). Dropping the $\bar{K}^0$ excitation, see remark below Eq.\ (\ref{bose}), the only relevant 
kaon dispersion relation is \cite{Alford:2007qa}
\be \label{excite}
\epsilon_k=\left\{ \begin{array}{cc} \left\{v_\pi^2k^2+M^2(T)+\mu^2-\sqrt{4\mu^2v_\pi^2k^2+[M^2(T)+\mu^2]^2}\right\}^{1/2} & 
\mbox{for}\; T<T_c \, ,\\ & \\
\sqrt{v_\pi^2k^2+M^2(T)}-\mu & \mbox{for}\; T>T_c \, , \end{array}\right.
\ee
with $M(T)$ determined from the stationarity equations (\ref{stationarity}). We have eliminated $\phi$ from the expression for 
$\epsilon_k$ by using Eq.\ (\ref{stat1}).
The critical temperature of the second-order phase transition for kaon condensation is denoted by $T_c$,
i.e., for temperatures below (above) $T_c$ we have $\phi\neq 0$ ($\phi=0$). The value of $T_c$ is estimated to be at least of the order of
tens of MeV \cite{Alford:2007qa}. This shows that for all temperatures relevant for compact stars, the kaons will be condensed 
(provided the parameters $m$
and $\mu$ are such that there is condensation at zero temperature). The self-consistent treatment ensures that for all temperatures
$M(T)>\mu$, and thus the energies given in Eq.\ (\ref{excite}) are real and positive as they should be.  Moreover, from Eq.\ (\ref{excite})
we see that $\epsilon_{k=0}=0$ for $T<T_c$. This shows that there is a massless Goldstone mode associated with kaon 
condensation.\footnote{We shall in the following use the term $K^0$ for this Goldstone mode in the CFL-$K^0$ phase 
as well as for the member of the meson octet arising from chiral symmetry breaking in the CFL phase. This is the most convenient 
terminology but one should keep in mind that these two excitations have very different dispersion relations.} 
It is the presence of this gapless mode that causes thermodynamic and
transport properties of the CFL-$K^0$ phase to differ from those of the
CFL phase. Strictly speaking, this Goldstone mode also has a small energy gap. This energy gap does not show up in 
our treatment within the effective theory given by the Lagrangian (\ref{Leff}). The energy gap is rather induced by the weak interactions 
and can be estimated to be in the keV range \cite{Son:2001xd}. Most of the temperature scales we show in this paper do 
not include temperatures below this range. Therefore, we shall neglect this effect and speak of an exact Goldstone mode with 
massless dispersion relation given by Eq.\ (\ref{excite}).

It is instructive to expand the kaon dispersion below $T_c$ for small momenta,
\be \label{lowk}
\epsilon_k^2 \simeq \frac{M^2(T)-\mu^2}{M^2(T)+\mu^2} v_\pi^2k^2 + \frac{2\mu^4}{[M^2(T)+\mu^2]^3} v_\pi^4k^4 \, .      
\ee
For small momenta, the dispersion is linear in $k$. However, for $T\to T_c$, the coefficient in front of the linear term goes to zero since 
$M(T\to T_c)\to\mu$. Consequently, around $T_c$ the quadratic part of the dispersion becomes important.

\subsection{One-loop approximation for the superfluid mode $H$}

\begin{figure} [t]
\begin{center}
\includegraphics[width=0.4\textwidth]{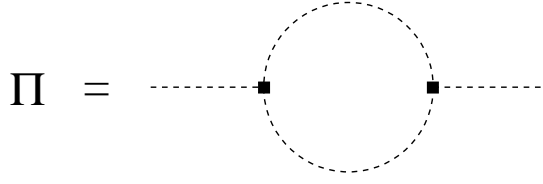}
\caption{
Diagrammatic representation of the one-loop self-energy $\Pi$ for the superfluid mode $H$. The lines represent the tree-level $H$ propagator $D_0$.
The vertex, here denoted by a square, is given in Eq.\ (\ref{Hvertex}).}
\label{figHself}
\end{center}
\end{figure}

Next we turn to the $H$ boson which is associated with the spontaneous breaking of $U(1)_B$. This Goldstone boson 
can be described by the effective Lagrangian \cite{Son:1999cm,Son:2000tu,Son:2002zn,Manuel:2004iv}
\be
{\cal L} = \frac{1}{2}(\partial_0\varphi)^2-\frac{v_H^2}{2}(\nabla\varphi)^2-
\frac{\pi}{9\mu_q^2}\partial_0\varphi \partial_\mu\varphi\partial^\mu\varphi
+\frac{\pi^2}{108\mu_q^4}(\partial_\mu\varphi\partial^\mu\varphi)^2\, ,
\ee
where $v_H=1/\sqrt{3}$. For notational convenience, we shall from now on abbreviate both $v_H$ and $v_\pi$ by $v$,
\be
v\equiv v_H=v_\pi \, .
\ee
For the calculation of the bulk viscosity we need the one-loop self-energy of the $H$,  
given by the diagram in Fig.\ \ref{figHself}, 
\be
\Pi(P) = \frac{4\pi^2}{81\mu_q^4}\frac{T}{V}\sum_K F(P,K)D_0(K)D_0(P-K) \, , 
\ee
where the vertex (squared) is    
\be\label{Hvertex}
F(P,K)\equiv [p_0(2K\cdot P-K^2)+k_0(P^2-2K\cdot P)]^2 \, , 
\ee
and the inverse tree-level propagator is $D_0^{-1}(K)=k_0^2-v^2 k^2$.
We need the imaginary part of the retarded self-energy \cite{Manuel:2004iv}
\bea
{\rm Im}\,\Pi(P) &=& \frac{4\pi^3}{81\mu_q^4} \frac{1}{f(p_0)}\sum_{e_1,e_2=\pm}\int\frac{d^3{\bf k}}{(2\pi)^3}\frac{e_1e_2}{4v^2k|{\bf p}-{\bf k}|}
F(e_1vk)f(e_1vk)f(e_2v|{\bf p}-{\bf k}|)  \delta(p_0-e_1vk-e_2v|{\bf p}-{\bf k}|) \, , 
\eea
where $F(e_1vk)\equiv F(P,K)|_{k_0=e_1vk}$.
The angular integration can be performed exactly. One finds that 
the self-energy assumes different forms depending on the sign of $p_0-vp$, 
\be \label{ImPi}
{\rm Im}\,\Pi(P)= \Pi^+(P)\Theta(p_0-vp)+\Pi^-(P)\Theta(vp-p_0) \, , 
\ee
where
\begin{subequations} \label{ImPipm}
\bea
\Pi^+(P)&\equiv& \frac{\pi}{324\mu_q^4v^7}\frac{p_0^2}{p}\frac{1}{f(p_0)}\int_{\frac{p_0-vp}{2v}}^{\frac{p_0+vp}{2v}}dk\, 
g^2(P,k)f(vk)f(p_0-vk) \, , \\
\Pi^-(P)&\equiv& \frac{2\pi}{324\mu_q^4v^7}\frac{p_0^2}{p}\frac{1}{f(p_0)}\int_{\frac{vp+p_0}{2v}}^\infty dk\, g^2(P,k)f(vk)[1+f(vk-p_0)] \, .
\eea
\end{subequations}
Here we have abbreviated 
\be
g(P,k)\equiv p_0^2-v^2p^2- 3(1-v^2)vk(p_0- vk) \, .
\ee
Denoting the one-loop $H$ propagator by
\be \label{oneloopH}
D^{-1}(P)=D^{-1}_0(P)+\Pi(P) \, , 
\ee
and neglecting the real part of $\Pi$ we have an approximate form of the imaginary part of the propagator,  
\be \label{ImD}
{\rm Im}\,D(P) \simeq \frac{{\rm Im}\,\Pi(P)}{(p_0^2-v^2p^2)^2+{\rm Im}\,\Pi^2(P)}\, ,
\ee
which we will need in the calculation of the bulk viscosity in Sec.\ \ref{bulk}. 

\subsection{Specific heat}
\label{specific}

The self-consistent formalism from Ref.\ \cite{Alford:2007qa}, summarized above, provides us with the tools 
to compute thermodynamical quantities of the CFL-$K^0$ phase. As a first application, we shall discuss the calculation of the specific heat. 
This is of physical relevance 
for instance for the cooling behavior of compact stars; see for instance Refs.\ \cite{Schafer:2004jp,Schmitt:2005wg,Alford:2007xm} 
for the specific heat of other quark matter phases. 
We will not need the result for the calculation of the bulk viscosity. However, the 
calculation shows in an exemplary way how to compute thermodynamical quantities in the 2PI formalism and will then in Sec.\ \ref{density}
be applied to the calculation of the kaon susceptibility. 

\begin{figure} [t]
\begin{center}
\hspace{2cm} Physical case ($K^0$-dominated)\hspace{3.5cm} Isospin-symmetric scenario ($K^0$ and $K^+$)
\hbox{
\includegraphics[width=0.45\textwidth]{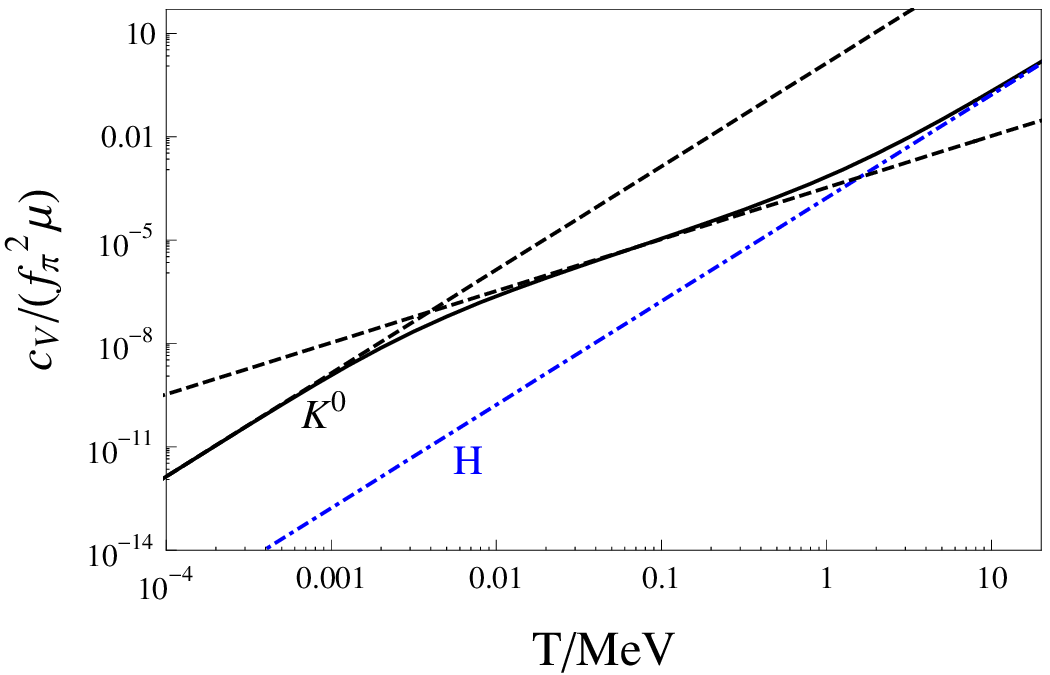}
\hspace{0.5cm}\includegraphics[width=0.45\textwidth]{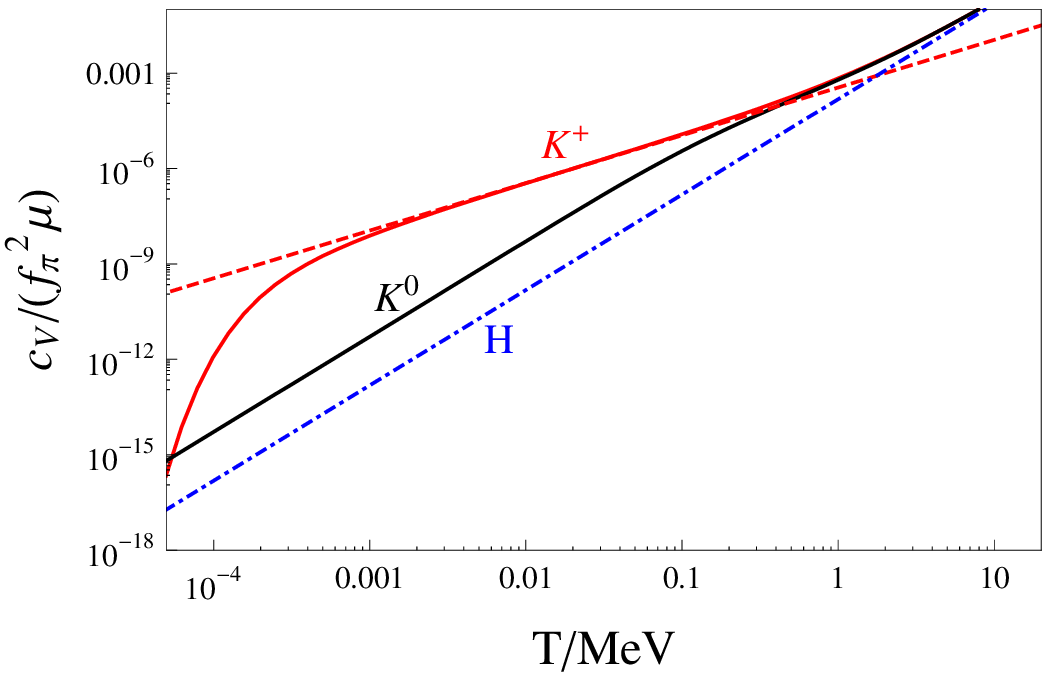}
}
\caption{
(Color online) Various contributions to the specific heat as a function of temperature in the CFL-$K^0$ phase. 
In both
panels we show the specific heat of the $K^0$ mesons as a solid black line, and the low-temperature specific heat
of the $H$ mesons as a dashed-dotted (blue) line. These
coincide at large $T$ because the low-$T$ approximation for the $H$ is identical to the large-$T$ approximation for the
$K^0$, see Eq.\ (\ref{cVH}) and Table \ref{tablecV}.
In the left panel we show
low-temperature approximations for the contributions of the linear (steeper dashed line, Eq.\ (\ref{cVlin})) 
and quadratic (shallower dashed line, Eq.\ (\ref{cVquad})) parts of the kaon dispersion. These are good approximations
in different temperature ranges.
In the right panel we show the contribution of the
$K^+$ mode for the case of an almost exact isospin symmetry (curved solid (red) line) and
the approximation from Eq.~(\ref{cVKp}) (straight dashed (red) line).
Parameter values: for the  
left panel we have used a kaon mass $m=4\,{\rm MeV}$, and a kaon chemical potential $\mu=4.01\,{\rm MeV}$; 
for the right panel we have used $m=4\,{\rm MeV}$, $\mu=4.501\,{\rm MeV}$ for the neutral kaons and 
and $m_{K^+}=4\,{\rm MeV}$, $\mu_{K^+}=4.5\,{\rm MeV}$ for the charged kaons. These 
parameters lead to critical temperatures of $T_c\simeq 8\,{\rm MeV}$ (left) and $T_c\simeq 39\,{\rm MeV}$ (right).}  
\label{figcV}
\end{center}
\end{figure}

The definition of the specific heat at constant volume $c_V$ is 
\be \label{defcV}
c_V = T\frac{\partial s}{\partial T} \, , \qquad s = \frac{\partial P}{\partial T} \, ,
\ee
where $s$ is the entropy density. 
It seems straightforward to take the second derivative of the pressure in Eq.\ (\ref{pressure}) with respect
to the temperature. However, the self-consistent treatment complicates this procedure. Since the pressure is a 
function of the self-consistent quantities $\phi$ and $M$, the implicit dependence of $\phi$ and $M$ on the temperature and the 
constraint of the stationarity equations have to be taken into account. We present the details of the calculation in Appendix \ref{AppA}.
The result is 
\be \label{cV}
c_V = \frac{1}{2}\int\frac{d^3{\bf k}}{(2\pi)^3}\frac{\epsilon_k^2}{T^2}\frac{1}{\cosh\frac{\epsilon_k}{T}-1}
-\frac{T}{2}\frac{\partial I}{\partial T}\frac{\partial M^2}{\partial T} \, ,
\ee
where $I$ is the integral defined in Eq.\ (\ref{defI}), and $\partial M^2/\partial T$ is given in Eq.\ (\ref{dMdT}). 
From this expression one obtains the discontinuity $\Delta c_V$ of the specific heat at the critical point,   
\be \label{DeltacV}
\Delta c_V=2T\left(\frac{\partial I}{\partial T}\right)^2
\frac{1}{1-4\alpha^2\left(\frac{\partial I}{\partial M^2}\right)^2} \, .
\ee
For all interesting parameters, the first term on the right-hand side of Eq.\ (\ref{cV}) is dominant and $\Delta c_V$ is small, as we shall 
see from the numerical results.
For small temperatures only small momenta of the kaons contribute. Therefore we can use the small-momentum kaon dispersion from 
Eq.\ (\ref{lowk}) to approximate the specific heat at small temperatures. The linear part of the dispersion produces 
a contribution cubic in the temperature,
\be \label{cVlin}
c_V^{\rm lin}\simeq \frac{2\pi^2T^3}{15v^3}\left(\frac{3\mu^2-m^2}{\mu^2-m^2}\right)^{3/2} \, ,
\ee
where we have approximated the thermal mass by its zero-temperature value, $M^2\simeq 2\mu^2-m^2$. The dispersion 
(\ref{lowk}) shows that at the critical temperature the linear term vanishes and the dispersion becomes quadratic in the momentum. 
In this case, the contribution to $c_V$ goes like $T^{3/2}$,
\be \label{cVquad}
c_V^{\rm quad}\simeq \frac{a\,T^{3/2}}{2^{11/4}\pi^2v^3}\frac{(3\mu^2-m^2)^{9/4}}{\mu^3} \, , \qquad a\equiv 
\int_0^\infty dx\frac{x^6}{\cosh x^2-1} \simeq 4.46 \, .
\ee
The kaon contribution has to be compared to the $H$ contribution. Approximating the dispersion by the linear
behavior $vk$, cf.\ the inverse tree-level propagator below Eq.\ (\ref{Hvertex}), we find for low temperatures
\be \label{cVH}
c_V^H\simeq \frac{2\pi^2T^3}{15v^3} \, .
\ee
By comparing this expression with Eq.\ (\ref{cVlin}) we conclude that the low-temperature contribution of the $K^0$ has the same $T$ 
dependence as the $H$ contribution, but that the $K^0$ contribution is always larger by a numerical factor which depends on the
zero-temperature kaon mass and chemical potential. We show the full numerical result of the kaon specific heat 
and its comparison with $c_V^{\rm lin}$, $c_V^{\rm quad}$, and $c_V^H$ in the left panel of Fig.\ \ref{figcV}. 
We see that there is 
a significant temperature regime in which $c_V^{\rm quad}$ is a good approximation to the specific heat, i.e., $c_V$ behaves as $T^{3/2}$ as opposed
to $T^3$. The presence of this temperature regime depends on the parameters. Here we have chosen the kaon mass and chemical potential 
such that this regime is relatively large ($\mu$ only slightly larger than $m$). For larger values of $\mu$, this regime 
is typically smaller or hardly visible (see right panel of Fig.\ \ref{figcV}). 
Note also that our temperature scale extends to very small temperatures in the keV regime. In this
regime we expect the small mass of the kaon, which is not present in our formalism, to play a role, see remark below Eq.\ (\ref{excite}).

The discontinuity at the critical point is small and not visible in the plot. At large temperatures, the specific heat 
again goes like $T^3$, however with a different prefactor than at low temperatures. In Table \ref{tablecV} we have listed the 
low-$T$ and high-$T$ behavior for the kaon specific heat together with the density and susceptibility, to be computed in Sec.\ \ref{density}.
In the table we also show the behavior for the case of no kaon condensation, i.e., for $m>\mu$. 

\begin{table*}[t]
\begin{tabular}{|c||c|c|c|} 
\hline
 & \multicolumn{2}{c}{low $T$}\vline  & high $T$ \\ \hline\hline
 & CFL-$K^0$ & CFL   & \;\; CFL \;\;\\ \hline\hline
\rule[-1.5ex]{0em}{6ex} kaon specific heat $c_V$ & $\;\;\displaystyle{\frac{2\pi^2}{15v^3}\left(\frac{m}{|\delta m|}\right)^{3/2}\, T^3}\;\;$ & 
$\displaystyle{\;\;\frac{\delta m^2m^{3/2}}{2\sqrt{2}\pi^{3/2}v^3T^{1/2}}\,e^{-\delta m/T}\;\;}$& 
$\displaystyle{\frac{2\pi^2}{15v^3}\,T^3}$\\[2ex] \hline
\rule[-1.5ex]{0em}{6ex}kaon density $n$ & $\displaystyle{\;\;4\mu f_\pi^2\frac{|\delta m|}{m}}$ & 
$\displaystyle{\frac{m^{3/2}T^{3/2}}{2\sqrt{2}\pi^{3/2}v^3}\,e^{-\delta m/T}}$
& $\displaystyle{\;\;\frac{\zeta(3)}{\pi^3v^3}\,T^3\;\;}$\\[2ex] \hline 
\rule[-1.5ex]{0em}{6ex}\;\; kaon susceptibility $\chi\;\;$ & $\displaystyle{\;\;4f_\pi^2\;\;}$ & 
$\displaystyle{\;\;\frac{m^{3/2}T^{1/2}}{2\sqrt{2}\pi^{3/2}v^3}\,e^{-\delta m/T}\;\;}$& $\displaystyle{\frac{1}{3v^3}\,T^2}$\\[2ex] \hline
\end{tabular}
\caption{Collection of thermodynamic properties of neutral kaons in the CFL-$K^0$ and CFL phases. Here ``high $T$'' 
means in particular $T>T_c$, i.e., there is no CFL-$K^0$ phase at these temperatures.
The zero-temperature kaon mass is denoted by $m$, and $\delta m\equiv m-\mu$ with the kaon chemical potential $\mu$.
In the CFL-$K^0$ phase we have $\delta m <0$, and we show the low-$T$ results to lowest order in $\delta m/m$ which corresponds to small condensates
$\phi/f_\pi\ll 1$. In the CFL phase, $\delta m>0$.} 
\label{tablecV}
\end{table*}

So far we have ignored the charged kaons. Due to the electric charge neutrality condition their number density is suppressed
at small temperatures \cite{Alford:2007qa}. However, in the isospin symmetric case they give rise to an additional Goldstone mode 
\cite{Miransky:2001tw,Schafer:2001bq,Alford:2007qa} which in principle gives a large contribution to the specific heat. To illustrate this 
contribution we ignore the neutrality constraint for simplicity and compute the specific heat for a fixed charged kaon chemical 
potential $\mu_{K^+}$. For low momenta and exact isospin 
symmetry $\mu=\mu_{K^+}$, $m=m_{K^+}$ the additional Goldstone mode has a 
quadratic dispersion, $\epsilon_k=v^2k^2/(2\mu^2)$. Consequently, its contribution to the specific heat goes like
$T^{3/2}$, just as the contribution from Eq.\ (\ref{cVquad}),
\be \label{cVKp}
c_V^{K^+} \simeq \frac{a\, T^{3/2}}{\sqrt{2}\pi^2 v^3}\mu_{K^+}^{3/2} \, .
\ee
We may use the formalism of Ref.\ \cite{Alford:2007qa}
to compute the thermal masses in the two-component system of neutral and charged kaons and derive the specific heat 
analogously to the above one-component system. The numerical result is shown in the right panel of Fig.\ \ref{figcV}. For this plot we have 
chosen the kaon masses and chemical potentials such that the isospin symmetry is almost exact. We see that there is a large contribution
from the charged kaon mode which disappears for sufficiently small temperatures. This is due to the small isospin symmetry violation in the 
parameters and hence a small charged kaon energy gap.

\section{Bulk viscosity in the CFL-$K^0$ phase}
\label{bulk}

We now turn to the main part of the paper where we use the results of the previous sections to compute the bulk viscosity in the CFL-$K^0$ phase.

\subsection{Definition of bulk viscosity}
\label{definition}

The definition of the bulk viscosity and the derivation of its expression in terms of the kaon rate can be found in Ref.\ \cite{Alford:2007rw}.
Therefore, here we only briefly summarize the most important relations and the underlying physics. In general, a relativistic superfluid
may have more than one bulk viscosity, related to stresses in the superfluid flow with respect to the normal flow 
\cite{Andersson:2006nr,Gusakov:2007px,Mannarelli:2008jq}. Here we 
neglect this effect and compute the bulk viscosity related to the normal fluid.
 
We are interested in a system with volume $V_0$ which undergoes a volume oscillation with amplitude $\delta V_0\ll V_0$ and frequency
$\omega$,
\be \label{volume}
V(t) = V_0 + \delta V_0\cos\omega t \, .
\ee
In the astrophysical setting, the oscillations are local volume oscillations with an (inverse) time scale typically of the order of 
the rotation frequency of the star which can be as large as $\omega/(2\pi)\sim 1\,{\rm ms}^{-1}$.  
The periodic change in volume induces a change in density. This change in density, in turn, may induce a change in the chemical composition 
of the matter. In CFL-$K^0$ matter, there will be a change in the strangeness content, described by an induced
nonzero $\delta\mu$, i.e., the equilibrium kaon chemical potential $\mu$ is shifted to the nonequilibrium value 
$\mu+\delta\mu$. The response of the system is to create (or annihilate) kaons in order to reequilibrate. The dominant processes 
considered here that change kaon number (and thus strangeness) are 
\begin{subequations} \label{KtoH}
\bea
K^0\leftrightarrow H+H \, , \label{KtoHH} \\
K^0+H\leftrightarrow H \, . \label{KHtoH}
\eea
\end{subequations}
These processes have also been considered in Ref.\ \cite{Alford:2007rw} for the case of uncondensed kaons. 
If the kaons are condensed there is also a cubic interaction,
induced by attaching one leg of the quartic interaction to the
condensate, which directly moves strangeness between the condensate
and the thermal kaon gas.  Here we neglect these processes since
they are proportional to the condensate. This is consistent with our
2PI treatment which is only valid for small condensates,
i.e.~$(\mu-m)/m\ll 1$.

The processes (\ref{KtoH}) induce, in response to the external oscillation $V(t)$, an oscillation in the kaon chemical potential $\mu(t)$. If 
the external oscillation and the system's response are out of phase, there will be dissipation, 
resulting in a nonzero value of the bulk viscosity. 
The bulk viscosity is maximized if the external oscillation and the rate of kaon production (annihilation) is on the same time scale.  
In this sense, bulk viscosity is a resonance phenomenon. In fact, it is the exact analogue of an electric circuit with alternating 
voltage that responds by an induced alternating current \cite{Alford:2006gy}.   

The definition of the bulk viscosity is
\be \label{defbulk}
\zeta = 2\frac{V_0^2}{\omega^2\delta V_0^2}\frac{dE}{dt} \, ,
\ee
where the dissipated power is given by 
\be \label{dEdt}
\frac{dE}{dt} = -\frac{1}{2}\frac{\delta V_0}{V_0}\omega{\rm Im}\,\delta P \, .
\ee
The imaginary part of the complex amplitude $\delta P$ is given by 
\be \label{ImP}
{\rm Im}\,\delta P = n\,{\rm Im}\,\delta \mu+  n_q {\rm Im}\,\delta \mu_q \, , 
\ee
where the complex amplitudes $\delta\mu$, $\delta\mu_q$ account for the oscillating change in the chemical potentials,
\begin{subequations} \label{dmut}
\bea
\mu(t) &=& \mu+ {\rm Re}(\delta \mu \,e^{i\omega t}) \, , \label{mut}\\
\mu_q(t) &=& \mu_q + {\rm Re}(\delta\mu_q  e^{i\omega t}) \, .  \label{mut2}
\eea
\end{subequations}
Here, $n_q$ and $\mu_q$ are the quark number density and chemical potential, while $n$ and $\mu$ are the corresponding kaon quantities.
Note that the effective potential (\ref{Veff}) also depends on the quark chemical potential $\mu_q$. 
In particular, the quark density is nonvanishing and there is an induced oscillation also in $\mu_q$. 
This effect, however, is small. In Appendix \ref{AppB} we present the general derivation of the bulk 
viscosity, taking into account the quark density effect. Here we proceed by only keeping the kaon terms. 
In this case, there is a single differential equation for $\delta\mu$,
\bea \label{diff}
\frac{\partial n}{\partial\mu}\frac{\partial \mu}{\partial t}&=&
\frac{\partial n}{\partial V}\frac{\partial V}{\partial t}+\Gamma_{K^0} \, .
\eea
The left-hand side of this equation simply means that the kaon density is a function of the kaon chemical potential, hence the time dependence
of $n$ can be expressed in terms of the time dependence of $\mu$. The
right-hand side expresses this change in terms of the volume change (first term) and the kaon rate (second term). The kaon rate $\Gamma_{K^0}$
is defined as the change in kaon number per time and volume due to the processes (\ref{KtoH}). For the microscopic 
definition of $\Gamma_{K^0}$ see Eq.\ (\ref{Gamma}) in Sec.\ \ref{rates}.  For sufficiently small $\delta\mu$ we can approximate $\Gamma_{K^0}$ 
by 
\be \label{lambda}
\Gamma_{K^0}\simeq -\lambda \,{\rm Re}(\delta\mu\, e^{i\omega t}) \, .
\ee
This defines the factor $\lambda$ to be positive (for $\delta\mu>0$, the system responds by annihilating kaons, hence $\Gamma_{K^0}<0$; 
on the other hand, if $\delta\mu<0$, the system responds by creating kaons and thus $\Gamma_{K^0}>0$; in both cases we thus have
$\lambda>0$). Inserting Eqs.\ (\ref{volume}), (\ref{mut}), and (\ref{lambda}) into Eq.\ (\ref{diff}) and inserting the resulting solution for 
${\rm Im}\,\delta \mu$ into Eqs.\ (\ref{ImP}), (\ref{dEdt}), and (\ref{defbulk}) yields the bulk viscosity
\be \label{bulkfinal}
\zeta = \frac{n^2}{\chi^2}\frac{\lambda}{\omega^2+(\lambda/\chi)^2} \, ,
\ee
where we have denoted the kaon number susceptibility
\be \label{chi}
\chi\equiv \frac{\partial n}{\partial \mu} \, .
\ee
In the following we shall evaluate the bulk viscosity (\ref{bulkfinal}). To this end, we shall compute the equilibrium density $n$ and
susceptibility $\chi$ in Sec.\ \ref{density}, the kaon rate $\lambda$ in Sec.\ \ref{rates}, and put the results together in Sec.\ \ref{results}.

\subsection{Kaon density and kaon susceptibility}
\label{density}

\begin{figure} [t]
\begin{center}
\hbox{\includegraphics[width=0.45\textwidth]{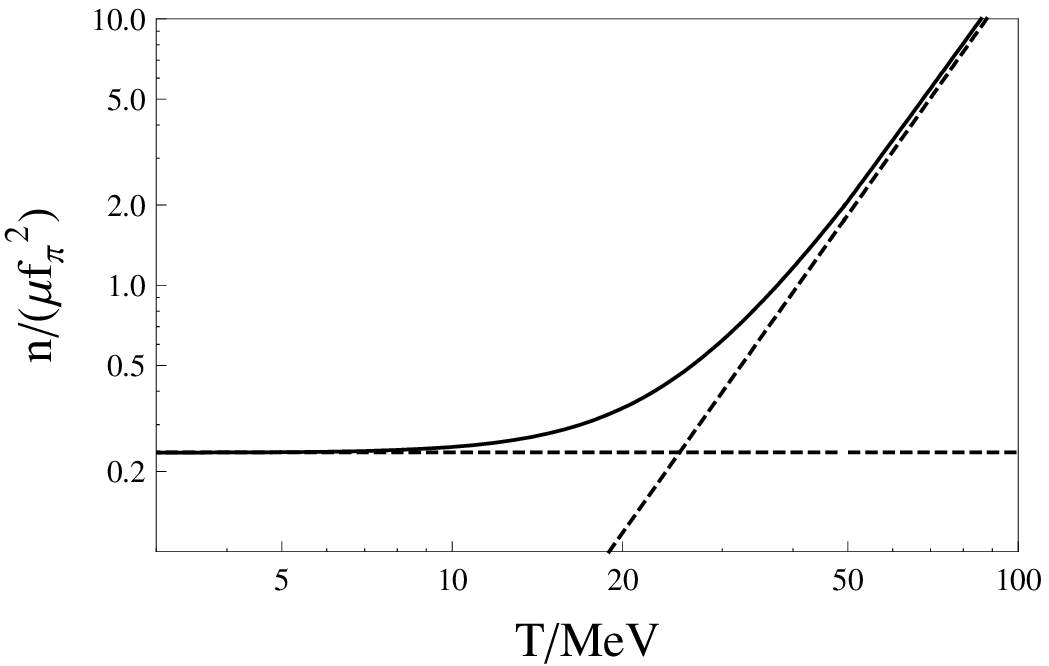}
\hspace{0.5cm}\includegraphics[width=0.45\textwidth]{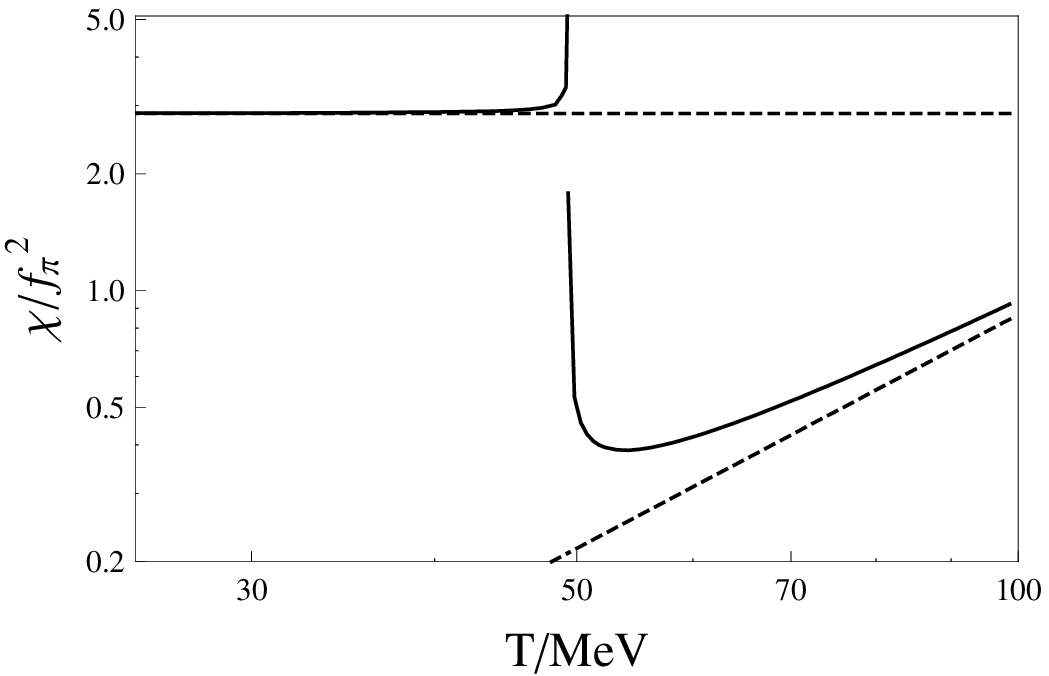}
}
\caption{
Kaon density (left panel) and kaon susceptibility (right panel) in the CFL-$K^0$ phase as a function of temperature (solid
lines: full numerical result; dashed lines: low-$T$ and high-$T$ approximations from Table \ref{tablecV}). Both quantities 
have nonzero values at $T=0$. 
We have used a kaon mass $m=4\,{\rm MeV}$, and a kaon chemical potential $\mu=4.3\,{\rm MeV}$. This leads to a critical temperature
$T_c\simeq 49\,{\rm MeV}$. At $T=T_c$, the density has a kink (not visible on the plotted scale), and the susceptibility is discontinuous. 
}
\label{figsusc}
\end{center}
\end{figure}

The kaon density is given by the negative of the derivative of the effective potential (\ref{Veff}) with respect to the 
chemical potential,
\be \label{density1}
n = -\frac{\partial U}{\partial \mu}-\frac{1}{2}{\rm Tr}\left[\frac{\partial S_0^{-1}}{\partial \mu}
S\right] \, . 
\ee
Here we have only taken the explicit derivative with respect to $\mu$. Implicit dependencies through the self-consistent 
condensate $\phi$ and the kaon mass
$M$ drop out of the density when we take the value of $n$ at the stationary point. Inserting the tree-level potential (\ref{Uphi}) and 
the propagators (\ref{S0}) and (\ref{fullprop}) yields (see also Ref.\ \cite{Alford:2007qa})
\be \label{n}
n \simeq \mu\phi^2\left[1-
\frac{1}{\mu}\frac{\partial \alpha}{\partial\mu}\left(\frac{\phi^2}{4}+I\right)\right] + \int\frac{d^3k}{(2\pi)^3}
f(\epsilon_k) \, ,
\ee
with the integral $I$ defined in Eq.\ (\ref{defI}) and the kaon excitation energy $\epsilon_k$ from Eq.\ (\ref{excite}). At zero temperature,
the integral over the Bose distribution function (as well as the integral $I$) vanishes, and 
the density is nonzero only because of a nonzero value of $\phi$; in other words, all kaons are condensed. Note that the effective 
coupling $\alpha$, see Eq.\ (\ref{alpha}), depends on the kaon chemical potential. With the zero-temperature value of the condensate 
$\phi^2=(\mu^2-m^2)/\alpha$
we easily obtain the low-temperature limit of the density, 
\be
n\simeq 6\mu f_\pi^2\frac{\left(\mu^2-m^2\right)\left(2\mu^2+m^2\right)}{\left(4\mu^2-m^2\right)^2} \, .
\ee
The expansion of this limit for small condensates and the opposite limit for high temperatures are shown Table \ref{tablecV}.
The two limits are compared with the full numerical result in the left panel of Fig.\ \ref{figsusc}.

The derivation of the susceptibility $\chi$ is more complicated. The susceptibility is the second derivative of the pressure
with respect to the chemical potential, and the dependence of $\phi$ and $M$ on $\mu$ does not drop out. 
Hence we have to compute $\chi$ analogously to the 
specific heat, which is also a second derivative of the 
pressure. Details of the calculation are presented in Appendix \ref{AppC}. The result is  
\bea\label{susc}
\chi&=& \left\{\begin{array}{cc} \chi_0 +\chi_1 +  \displaystyle{ \frac{1}{2T}\int\frac{d^3{\bf k}}{(2\pi)^3} \frac{1}{\cosh\frac{\epsilon_k}{T}-1}} & 
\mbox{for}\; T<T_c \, ,\\ & \\
\tilde{\chi}_1 + \displaystyle{\frac{1}{2T} \int\frac{d^3{\bf k}}{(2\pi)^3} \frac{1}{\cosh\frac{\epsilon_k}{T}-1}} & \mbox{for}\; T>T_c \, , \end{array}\right.
\eea
where $\chi_0$, $\chi_1$, and $\tilde{\chi}_1$ are given in Eqs.\ (\ref{chi0}) and (\ref{chi01}).
The zero-temperature limit is obtained from $\chi_0$ upon
approximating the condensate by its zero-temperature value, $\phi^2\simeq (\mu^2-m^2)/\alpha$,
\be
\chi\simeq 6f_\pi^2\frac{m^6+15m^4\mu^2-6m^2\mu^4+8\mu^6}{\left(4\mu^2-m^2\right)^3} \, .
\ee
The high-$T$ behavior is dominated by the momentum integral in Eq.\ (\ref{susc}). Table \ref{tablecV} shows both limit cases.
The full numerical result for the susceptibility and its comparison to the limit cases is shown in the right panel of Fig.\ \ref{figsusc}.
In contrast to the density and the specific heat we observe a strong discontinuity at the critical point. 

\subsection{Rates of the processes $K^0 \leftrightarrow H+H$ and $K^0 +H \leftrightarrow H$}
\label{rates}

\begin{figure} [t]
\begin{center}
\includegraphics[width=0.5\textwidth]{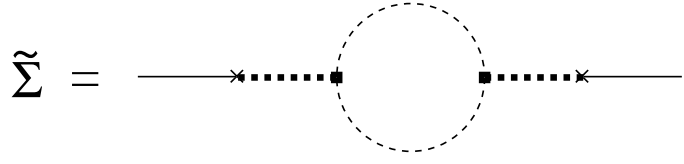}
\caption{
Kaon self-energy $\tilde{\Sigma}$ needed to compute the rate of the processes $K^0\leftrightarrow H+H$ and $K^0+H\leftrightarrow H$.
Thin dashed lines are tree-level $H$ propagators $D_0$, thick dashed lines are full $H$ propagators $D$, and solid lines are $K^0$ propagators. 
The $K^0\ H$ vertex, here denoted
by a cross, is $G_{ds}f_\pi f_H(p_0^2-v_{ds}^4p^2)$. The $H$ vertex, denoted by a square, is given in Eq.\ (\ref{Hvertex}).}
\label{figsigmaprime}
\end{center}
\end{figure}

The final ingredient to the bulk viscosity (\ref{bulkfinal}) is the rate $\Gamma_{K^0}$ due to the processes (\ref{KtoH}). 
To this end, we need the imaginary part of the kaon self-energy $\tilde{\Sigma}$ given by the diagram in Fig.\ \ref{figsigmaprime}. 
This is 
simply the imaginary part of the one-loop $H$ propagator (\ref{ImD}) multiplied by the square of the $K^0H$ vertex 
$G_{ds}f_\pi f_H(p_0^2-v_{ds}^4p^2)$ \cite{Alford:2007rw}, where 
\be \label{Gds}
G_{ds} \equiv -\sqrt{2}V_{ud}V_{us}G_F \, , \qquad v_{ds}=\frac{1}{\sqrt{3}} \, , \qquad f_H^2\equiv \frac{3\mu_q^2}{8\pi^2} \, , 
\ee
with the Fermi coupling $G_F$ and the entries of the CKM matrix $V_{ud}$, $V_{us}$. (Microscopically, the $K^0H$ interaction can be 
understood as the weak process $d+\bar{s}\leftrightarrow u+\bar{u}$, for details see Ref.\ \cite{Alford:2007rw}.) Consequently,
we obtain 
\be
{\rm Im}\,\tilde{\Sigma}(P) = \frac{G_{ds}^2f_\pi^2 f_H^2(p_0^2-v_{ds}^4p^2)^2{\rm Im}\,\Pi(P)}{(p_0^2-v^2p^2)^2+{\rm Im}\,\Pi^2(P)} \, .
\ee
We can now compute the kaon production rate $\Gamma_{K^0}$. In the 
closed-time path formalism (for calculations in a similar context using this formalism see for instance 
Refs.\ \cite{Schonhofen:1994zf,Schmitt:2005wg,Alford:2006gy,Jaikumar:2005hy}) the rate can be written as
\be \label{kinetic}
\Gamma_{K^0} = \frac{1}{4}\int\frac{d^4 P}{(2\pi)^4}[G^<(P)\tilde{\Sigma}^>(P)-G^>(P)\tilde{\Sigma}^<(P)] \, .
\ee
Here, the self-energies are given by  
\begin{subequations} \label{SigmaSigma}
\bea
\tilde{\Sigma}^>(P)&=&-2i[1+f(p_0)]\,{\rm Im}\,\tilde{\Sigma}(P) \, , \\
\tilde{\Sigma}^<(P)&=&-2if(p_0)\,{\rm Im}\,\tilde{\Sigma}(P) \, , 
\eea
\end{subequations}
and the propagators are
\begin{subequations} \label{GG}
\bea
G^>(P)&=&-\frac{2\pi i}{\epsilon_p'}[1+f(\epsilon_p')]\,\delta(p_0-\epsilon_p'-\delta\mu) \, , \\
G^<(P)&=&-\frac{2\pi i}{\epsilon_p'}f(\epsilon_p')\,\delta(p_0-\epsilon_p'-\delta\mu) \, . 
\eea
\end{subequations}
In these propagators we have introduced chemical nonequilibrium by a nonzero $\delta\mu$ which enters the kaon excitation
energy $\epsilon_p'\equiv \epsilon_p(\mu\to\mu+\delta\mu)$. In $\epsilon_p'$ the self-consistent mass $M$ is determined 
according to the modified chemical potential. This is necessary since the value of the
kaon mass $M$ is governed by the strong interaction and thus the mass adjusts itself instantaneously compared to the 
equilibration time of the weak processes (\ref{KtoH}). The energy $\epsilon_p'+\delta\mu$ in the $\delta$-functions of the 
propagators can be understood from the case without kaon condensation. In this case, we have $\epsilon_p=E_p-\mu$, and $\delta\mu$ 
simply cancels, $\epsilon_p'+\delta\mu=\epsilon_p$ (ignoring the change in the self-consistent mass). In the case with condensation, this 
cancellation is only partial.  
Inserting Eqs.\ (\ref{SigmaSigma}) and (\ref{GG}) into Eq.\ (\ref{kinetic})
yields
\bea \label{Gamma}
\Gamma_{K^0}&=& G_{ds}^2f_\pi^2 f_H^2\int\frac{d^4P}{(2\pi)^4}\frac{\pi}{\epsilon_p'}\delta(p_0-\epsilon_p'-\delta\mu)
\left\{[1+f(\epsilon_p')]f(p_0)-f(\epsilon_p')[1+f(p_0)]\right\}
\frac{(p_0^2-v_{ds}^4p^2)^2{\rm Im}\,\Pi(P)}{(p_0^2-v^2p^2)^2+{\rm Im}\,\Pi^2(P)} \, .
\eea
From this expression it is clear that $\delta\mu=0$ corresponds to chemical equilibrium because in this case forward and backward rate 
cancel each other, and $\Gamma_{K^0}=0$. For small $\delta\mu$ we may approximate
\be \label{dmuapprox}
[1+f(\epsilon_p')]f(\epsilon_p'+\delta\mu)-f(\epsilon_p')[1+f(\epsilon_p'+\delta\mu)]\simeq-f(\epsilon_p)[1+f(\epsilon_p)]\frac{\delta\mu}{T} \, .
\ee
This shows that the sign of the rate is determined by the sign of $\delta\mu$. 
For positive $\delta\mu$ the system reacts with annihilating 
kaons, and $\Gamma_{K^0}<0$, while for negative $\delta\mu$ kaons are created, and $\Gamma_{K^0}>0$.
With the approximation (\ref{dmuapprox}) we see that, to lowest order in $\delta\mu$, the rate is
\be \label{Gammaapprox}
\lambda=-\frac{\Gamma_{K^0}}{\delta\mu}\simeq \frac{G_{ds}^2f_\pi^2 f_H^2}{T}\int\frac{d^3{\bf p}}{2\epsilon_p(2\pi)^3}\,
f(\epsilon_p)[1+f(\epsilon_p)]\,
\frac{(\epsilon_p^2-v_{ds}^4p^2)^2{\rm Im}\Pi(\epsilon_p,{\bf p})}{(\epsilon_p^2-v^2p^2)^2+{\rm Im}\Pi^2(\epsilon_p,{\bf p})} \, ,
\ee
in agreement with Ref.\ \cite{Alford:2007rw}.
We shall use Eq.\ (\ref{Gammaapprox}) for the numerical evaluation of the
rate. Before we turn to the results let us first rewrite the rate into a more instructive form. Inserting the $H$ self-energy from 
Eqs.\ (\ref{ImPi}) and (\ref{ImPipm}) into Eq.\ (\ref{Gamma}), we obtain
\bea \label{rate2}
\Gamma_{K^0} &=& \frac{G_{ds}^2f_\pi^2 f_H^2}{648\pi\mu_q^4v^7}\int_0^\infty dp\,  
\frac{p \,(\epsilon_p'+\delta\mu)^2}{\epsilon_p'} \frac{[(\epsilon_p'+\delta\mu)^2-v_{ds}^4p^2]^2}
{[(\epsilon_p'+\delta\mu)^2-v^2p^2]^2+{\rm Im}\Pi^2(\epsilon_p'+\delta\mu,{\bf p})}\non
&& \times\, \Big\{ \Theta(\epsilon_p'+\delta\mu-vp)[1+f(\epsilon_p')]
\int_{\frac{\epsilon_p'+\delta\mu-vp}{2v}}^{\frac{\epsilon_p'+\delta\mu+vp}{2v}}
dk\,g^2(\epsilon_p'+\delta\mu,p,k)f(vk)f(\epsilon_p'+\delta\mu-vk) \non
&& +\,2\Theta(vp-\epsilon_p'-\delta\mu) [1+f(\epsilon_p')]\int_{\frac{\epsilon_p'+\delta\mu+vp}{2v}}^\infty 
dk\,g^2(\epsilon_p'+\delta\mu,p,k)f(vk)[1+f(vk-\epsilon_p'-\delta\mu)]  \non
&& -\, \Theta(\epsilon_p'+\delta\mu-vp)f(\epsilon_p')\int_{\frac{\epsilon_p'+\delta\mu-vp}{2v}}^{\frac{\epsilon_p'+\delta\mu+vp}{2v}}
dk\,g^2(\epsilon_p'+\delta\mu,p,k)
[1+f(vk)][1+f(\epsilon_p'+\delta\mu-vk)]\non
&& - \,2\Theta(vp-\epsilon_p'-\delta\mu) f(\epsilon_p')\int_{\frac{\epsilon_p'+\delta\mu+vp}{2v}}^\infty 
dk\,g^2(\epsilon_p'+\delta\mu,p,k)[1+f(vk)]f(vk-\epsilon_p'-\delta\mu) \Big\} \, .
\eea
This expression is useful to discuss the separate processes that contribute to the rate. From the structure of the Bose distribution 
functions (interpreting $f$ as an ingoing and $1+f$ as an outgoing particle) we see that  
the kaon rate is composed of four contributions: the processes $H+H\to K^0$ and $H\to K^0+H$ (first and second term, respectively) and the inverse 
processes  $K^0\to H+H$ and $H+K^0\to H$ (third and fourth term, respectively). As expected, we see that the kaon-creating processes 
yield a positive contribution to $\Gamma_{K^0}$ while the kaon-annihilating processes yield a negative contribution. Furthermore, we conclude
that in the CFL-$K^0$ phase (and for infinitesimal $\delta\mu$) only the processes $H\leftrightarrow K^0+H$ are allowed. This can be seen 
from the dispersion relation in 
Eq.\ (\ref{excite}) which implies $\epsilon_p<vp$ for $T<T_c$. Then, due to the step functions in Eq.\ (\ref{rate2}), the process 
$H+H\leftrightarrow K^0$ is excluded. This is of course a direct consequence of energy conservation.

For small temperatures we can derive an analytic expression for the rate in the CFL-$K^0$ phase. The derivation 
is presented in Appendix \ref{AppD}. The result is
\be \label{lambdasmallT}
\lambda_{{\rm CFL-}K^0} = 
\frac{G_{ds}^2f_\pi^2f_H^2}{2\pi}\sqrt{\frac{\mu^2-m^2}{3\mu^2-m^2}}\,\frac{m^4}{\mu^4}\,Q(m,\mu)\,\frac{T^7}{\mu_q^4} \, ,
\ee
where $Q(m,\mu)$ is a complicated dimensionless function of the zero-temperature mass $m$ and chemical potential $\mu$ given in 
Eq.\ (\ref{Q}). We can simplify this expression by defining 
\be
\delta m \equiv m-\mu
\ee
and expanding for small values of $|\delta m|/m$, which corresponds to 
small kaon condensates. This yields
\be \label{lambdasmallT1}
\lambda_{{\rm CFL-}K^0} = \frac{80\,G_{ds}^2f_\pi^2f_H^2}{\pi}\,\frac{T^7}{\mu_q^4}\,\left[1+{\cal O}\left(\frac{|\delta m|}{m}\right)\right] \, .
\ee
The full evaluation of the rate has to be done numerically; we have checked that the expression (\ref{lambdasmallT})
is in very good agreement with the full result up to temperatures $T\lesssim 1\,{\rm MeV}$. The additional approximation (\ref{lambdasmallT1})
is accurate to within 30\% for values of $|\delta m|/m \lesssim 0.05$. 

\begin{figure} [t]
\begin{center}
\hbox{\includegraphics[width=0.45\textwidth]{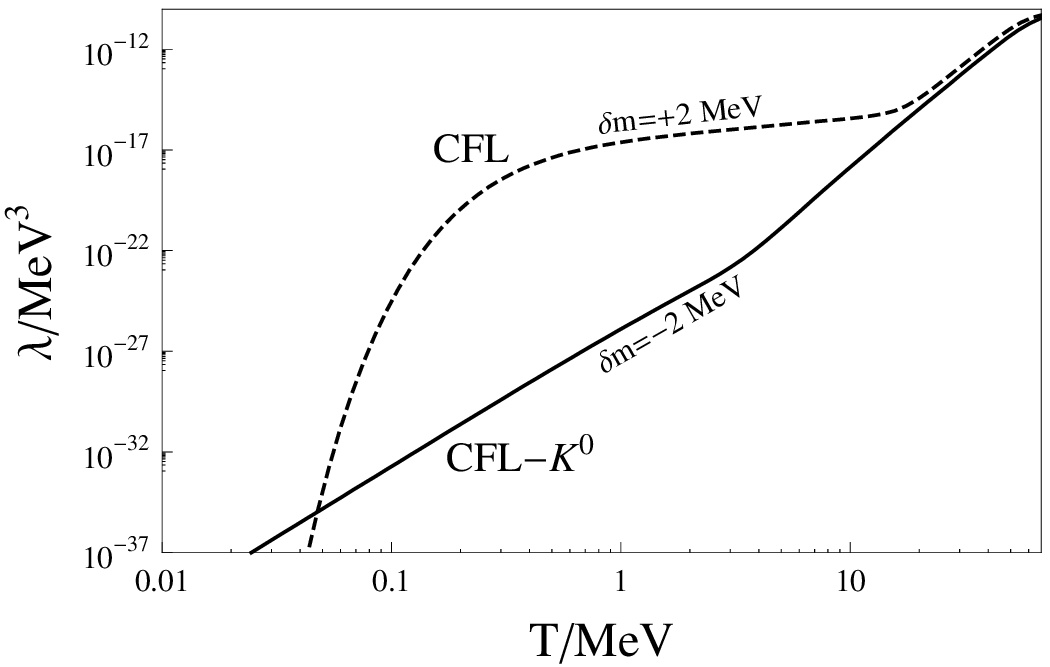}
\hspace{0.5cm}\includegraphics[width=0.45\textwidth]{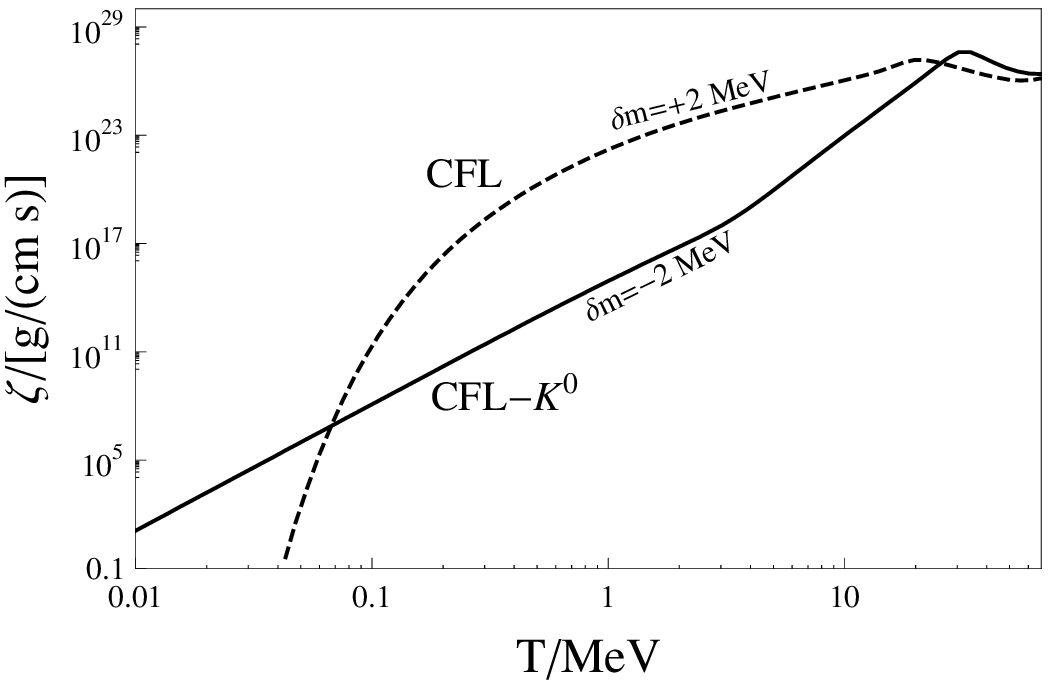}
}
\caption{Left panel: Kaon decay rate $\lambda=-\Gamma_{K^0}/\delta\mu$ as a function of temperature. We compare the rate in the CFL-$K^0$ phase 
(solid line) with the rate in the CFL phase (dashed line). We have set the kaon mass to $m=10\,{\rm MeV}$. The kaon chemical potentials 
are chosen slightly above, $\mu=12\,{\rm MeV}$, and slightly below, $\mu=8\,{\rm MeV}$, the kaon mass (i.e., $\delta m=-2\,{\rm MeV}$ 
for CFL-$K^0$, $\delta m=+2\,{\rm MeV}$ for CFL). The quark chemical potential 
is $\mu_q=400\,{\rm MeV}$. The critical temperature is located at the right boundary of the 
temperature scale, i.e, the entire solid curve describes a kaon-condensed system. Right panel: bulk viscosity $\zeta$ for the same parameters
and a compression frequency of $\omega/(2\pi)=1{\rm ms}^{-1}$.}
\label{figrate}
\end{center}
\end{figure}

In the left panel of Fig.\ \ref{figrate} we show the numerical result for $\lambda$ as a function of the temperature. 
From Eq.\ (\ref{Gds}) we have 
$G_{ds}\simeq - 7.8 \cdot 10^{-13} \,{\rm MeV}^{-2}$; moreover, we use $f_H\simeq f_\pi\simeq 100\,{\rm MeV}$. We compare the 
rate in the CFL-$K^0$ phase with the one in the CFL phase. 
To realize these two cases we choose the kaon chemical potential slightly 
above and slightly below the kaon mass. 

The main features of the two rates are as follows. At very small temperatures, the rate in the CFL-$K^0$ phase is larger than in the CFL phase;
parametrically, the former behaves as $T^7$ while the latter is exponentially suppressed. At very large temperatures, the two rates are 
almost identical.
In the intermediate temperature regime however, which is the regime relevant for neutron stars, the rate in the CFL phase is much larger. 
The reason is a larger phase space for the weak process, as we shall explain in detail in the following.

The kaon width is largest for kaons of momentum $p=\bar{p}$
such that the internal $H$ particle in the kaon self-energy diagram 
Fig.~\ref{figsigmaprime} is closest to being on shell: the
$H$ and $K^0$ with this momentum both have the same energy, 
$\epsilon_{p}=vp$, and the denominator of Eq.\ (\ref{Gammaapprox}) 
is then minimized. From the kaon excitation energies 
in Eq.\ (\ref{excite}) we conclude
\be \label{barp}
\bar{p}=\left\{\begin{array}{cc} \displaystyle{\frac{M^2-\mu^2}{2v\mu}}  & 
\rm{CFL} \, ,\\ & \\
0 & \;\;\mbox{CFL-}K^0 \, .  \end{array}\right. 
\ee
In the CFL-$K^0$ phase there will be much less phase space
associated with the maximum in the kaon width, because it occurs at zero momentum, 
so one expects the total kaon rate to be lower.
Whether the additional phase space in the CFL case is available and has a significant effect on the rate 
depends on the temperature. 
For sufficiently large temperatures, $T\gg \bar{p}$, where states well above $p=\bar{p}$ are thermally
populated, the effect of the peak at $p=\bar{p}$ is negligible.
Moreover, for large momenta the kaon dispersion is independent of whether there 
is a condensate or not; the dispersion then is simply linear, $\epsilon_p\simeq vp$. This is the reason why the two curves in 
the left panel of Fig.\ \ref{figrate} are, at large $T$, almost on top of each other (the small difference in the curves is simply due to the 
different values of $\mu$, not because of any qualitative difference between the phases). In other words, besides the values for 
$\bar{p}$ given in Eq.\ (\ref{barp}), $\epsilon_{p}=vp$ is also satisfied
asymptotically for large $p$ (in both CFL-$K^0$ and CFL phases alike). This momentum regime becomes available at large temperatures, 
and thus the rates become almost identical. For smaller (but not too small) temperatures the peak in the integrand in the CFL phase
is responsible for the large kaon rate compared to the CFL-$K^0$ phase. For sufficiently small temperatures, $T\ll \bar{p}$, states
around $p=\bar{p}$ are no longer populated. Therefore, the rate in the CFL-$K^0$ phase, which has its largest contribution from momenta 
close to zero, becomes larger than the rate in the CFL phase which becomes exponentially suppressed.

\subsection{Results for the bulk viscosity}
\label{results}

We can now put together the results from the previous subsections to compute the bulk viscosity. 
We insert the numerical results for the density $n$, see Eq.\ (\ref{n}) and left panel of Fig.\ \ref{figsusc}, for the 
susceptibility $\chi$, see Eq.\ (\ref{susc}) and right panel of Fig.\ \ref{figsusc}, and for the kaon rate $\lambda$, 
see Eq.\ (\ref{Gammaapprox}) and left panel of Fig.\ \ref{figrate}, into the expression for the bulk viscosity (\ref{bulkfinal}).
We shall first analyze the features of the bulk viscosity below the critical temperature and compare the result with the bulk viscosity
in the CFL phase. We shall use $\delta m = m -\mu$
as a parameter to distinguish between the CFL phase ($\delta m >0$) and the CFL-$K^0$ phase ($\delta m<0$).
Then, we analyze the behavior of the bulk viscosity at the critical point. This of academic rather
than physical interest since
the critical temperature is likely to exceed temperatures reached in a compact star. Finally, we shall discuss the parameter dependence 
on $\delta m$ and compare the results with the bulk viscosity in other quark matter phases.   

\subsubsection{CFL-$K^0$ vs.~CFL bulk viscosity}
   
We first use the parameters of the rates shown in the left panel of Fig.\ \ref{figrate} to compute the corresponding bulk viscosities 
of the CFL and CFL-$K^0$ phases. The result is shown in the right panel of Fig.\ \ref{figrate}. We have chosen an oscillation frequency
$\omega/(2\pi)=1\,{\rm ms}^{-1}$ which is typical for a compact star. Let us first discuss the gross features of the result, valid for both 
CFL and CFL-$K^0$ phases.
As explained in Sec.\ \ref{definition}, if the prefactor $n^2/\chi^2$ is held
constant, the bulk viscosity becomes maximal if the frequency $\omega$ and the 
rate of the processes (\ref{KtoH}) are on the same timescale.
More precisely, we have to compare $\omega$ with the effective rate 
$\lambda/\chi$ which appears in the bulk viscosity. 
As a very rough estimate, we read off from Fig.\ \ref{figsusc} that the susceptibility is of the order of 
$\chi\sim 10^4 \,{\rm MeV}^2$ (using $f_\pi\simeq 100\,{\rm MeV}$). 
This means that the kaon rate has to be of the order of
$\lambda\sim 10^{-13}\,{\rm MeV}^3$ to render the effective rate $\lambda/\chi$ of the order of a frequency in the ms$^{-1}$ regime.
We see from the left panel of Fig.\ \ref{figrate} that this is the case for large temperatures, $T\gtrsim 20\,{\rm MeV}$. And indeed 
we observe that the bulk viscosity has a maximum in this temperature regime. 
Far below these temperatures we have  
$\omega\gg\lambda/\chi$, and the bulk viscosity can be approximated by $\zeta\simeq \lambda n^2/(\chi^2 \omega^2)$. 
We now discuss the CFL-$K^0$ and CFL results separately in this 
low temperature regime. 

In the CFL-$K^0$ phase, the entire temperature dependence of the low-temperature bulk viscosity is given by the rate $\lambda$
because the density $n$ and the susceptibility $\chi$ tend to constant values at low $T$. 
This can also be seen by comparing the two
curves in the left and right panels of Fig.\ \ref{figrate}.
With 
the low-temperature expression for the rate (\ref{lambdasmallT1}) and the values for $n$ and $\chi$ from Table \ref{tablecV} 
we obtain the bulk viscosity in the CFL-$K^0$ phase at small temperatures. To lowest order in $|\delta m|/m$, corresponding
to small condensates, we find 
\be
\zeta_{{\rm CFL-}K^0}\simeq \frac{80\,G_{ds}^2f_\pi^2f_H^2}{\pi}\,\frac{\delta m^2\,T^7}{\omega^2\mu_q^4} \, .
\ee
In the CFL phase without kaon condensation, 
the rate behaves very differently at small temperatures: 
it is exponentially suppressed, $\lambda_{\rm CFL} \propto 
\exp(-\delta m/T)$. The prefactor $n^2/\chi^2$
contributes an additional factor of $T^2$, $\zeta_{\rm CFL} = T^2\lambda_{\rm CFL}/\omega^2 $.
In Fig.\ \ref{figrate} we confirm that the bulk viscosity at low $T$ in the CFL phase has a different 
$T$ dependence than the kaon rate.

\subsubsection{Critical behavior of the bulk viscosity}

Next we turn to the behavior of the bulk viscosity at the critical temperature
where kaon condensation disappears, and there is a transition from
the CFL-$K^0$ phase to CFL.
This is probably not relevant to astrophysics, for two reasons.
Firstly, one would have to fine tune the kaon mass and chemical potential to
bring this critical temperature down to typical compact star temperatures.
Secondly, the critical temperature for kaon condensation is
of the same order as the critical temperature for the underlying CFL condensate 
itself \cite{Alford:2007qa}, so there may really be a transition to unpaired
quark matter. However, from the theoretical point of view it might be interesting to 
compare the critical behavior of the bulk viscosity in the present context with the critical behavior in related systems. 
For example, the bulk viscosity in a hot quark-gluon plasma has recently
been studied \cite{Arnold:2006fz,Meyer:2007dy,Karsch:2007jc}, and simple
models not unlike the one we use for kaons in this paper have been employed to study the behavior at the critical point \cite{Chen:2007jq}, see
also \cite{Chen:2007kx}.

\begin{figure}[t]
\begin{center}
\includegraphics[width=0.55\textwidth]{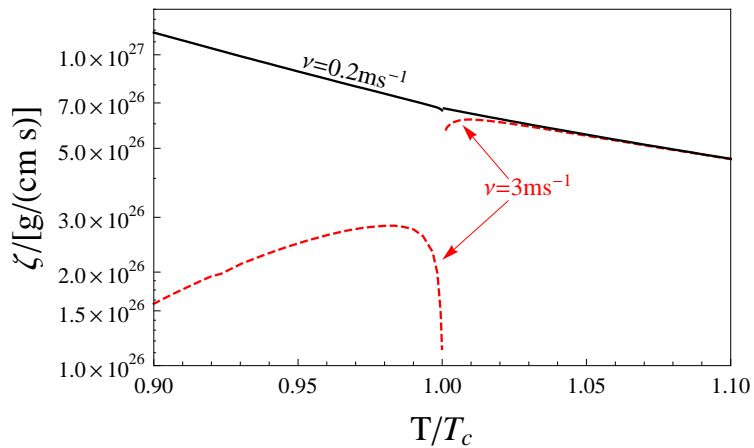}
\caption{(Color online) 
Behavior of the bulk viscosity around the critical temperature, here chosen to be $T_c\simeq 37\,{\rm MeV}$ ($m=10\,{\rm MeV}$, 
$\mu=10.5\,{\rm MeV}$), for different frequencies $\nu\equiv \omega/(2\pi)$. 
For $\nu\gtrsim 1$~ms$^{-1}$ the discontinuity becomes very noticeable.
}
\label{figbulkTc}
\end{center}
\end{figure}

With the help of Eq.\ (\ref{bulkfinal}) we can make
qualitative predictions of the behavior of the bulk viscosity at the critical point.
For small values of the frequency, $\omega\ll\lambda/\chi$, 
the viscosity behaves like $\zeta\simeq n^2/\lambda$. Both the density $n$ and the rate $\lambda$ are continuous 
at the phase transition, hence the bulk viscosity is continuous too. On the other hand, for large frequencies, $\omega\gg\lambda/\chi$, 
the bulk viscosity 
behaves like $\zeta\simeq n^2\lambda/(\chi^2\omega^2)$. From Fig.\ \ref{figsusc} we know that the susceptibility $\chi$ is discontinuous at 
$T_c$ (and large for temperatures below and close to $T_c$). Consequently, we expect the bulk viscosity to be discontinuous too
(and small slightly below $T_c$). We show the numerical result
for the bulk viscosity around the critical point in Fig.\ \ref{figbulkTc}. The case of an almost continuous behavior as well as the 
case of a strongly discontinuous behavior is shown (for the latter we have chosen a
frequency $\omega/(2\pi)=3\,{\rm ms}^{-1}$ which is larger than any known compact star rotation rate).

\subsubsection{Parameter dependence and comparison to other quark phases}

\begin{figure} [t]
\begin{center}
\includegraphics[width=0.55\textwidth]{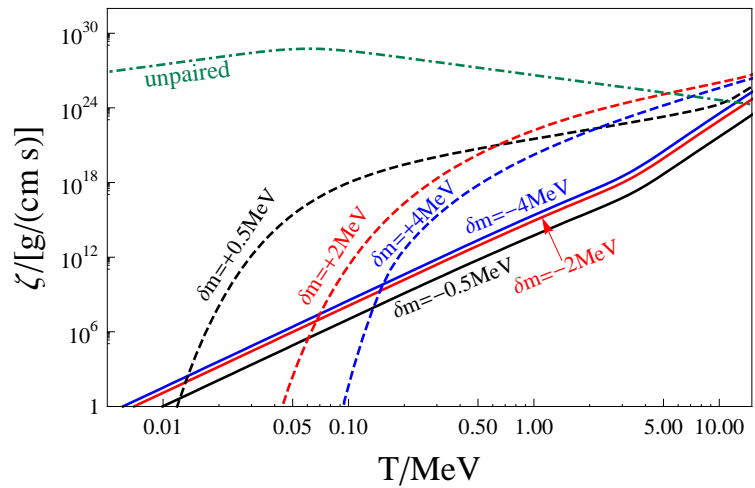}
\caption{(Color online) Bulk viscosity $\zeta$ due to strangeness equilibration for the CFL-$K^0$ phase (solid lines), 
the CFL phase (dashed lines), and unpaired quark matter 
(dashed-dotted line). For all curves, the quark chemical potential is $\mu_q =400\,{\rm MeV}$, the compression frequency 
$\omega/(2\pi)=1\,{\rm ms}^{-1}$,
and the kaon mass $m=10\, {\rm MeV}$. 
We have chosen different values of the kaon chemical potential $\mu$. Positive values of $\delta m= m-\mu$ 
correspond to the CFL phase (no kaon condensation) while negative values of $\delta m$ correspond to the CFL-$K^0$ phase
(kaon condensation).}
\label{figbulk}
\end{center}
\end{figure}

Finally, in Fig.\ \ref{figbulk}, we present the bulk viscosity for several values of the kaon chemical potential $\mu$ and compare the result 
with the bulk viscosity of unpaired quark matter \cite{Madsen:1992sx}.
As discussed in Sec.\ \ref{chiral}, the values of $\mu$ and the kaon mass $m$ are poorly known at densities
relevant for compact stars. We therefore vary the relevant parameters over ranges
of values that are plausible for conditions in a compact star.
We fix the kaon mass at $m=10\,{\rm MeV}$ and study six values of
the effective kaon chemical potential $\mu$; for three of them
$\delta m=m-\mu$ is positive (corresponding to the CFL phase)
and for the other three $\delta m$ is negative (corresponding to the CFL-$K^0$ phase). 
We have restricted our plot
to temperatures appropriate to compact stars, i.e., we have not shown temperatures larger than $15\,{\rm MeV}$.

In the CFL phase, the energy gap for the kaon is $\delta m$, so
the thermal population of kaons, and hence the  kaon-decay contribution to 
the bulk viscosity, will be very sensitive to the value of $\delta m$,
and will drop rapidly as $\exp(-\delta m/T)$ for $T\ll\delta m$.
In the CFL-$K^0$ phase, on the other hand, there is always a massless
Goldstone kaon, so the bulk viscosity due to kaon decay should be less sensitive
to $\delta m$, and should not drop exponentially at low temperatures.
These expectations are borne out in Fig.\ \ref{figbulk}.
Although the CFL-$K^0$ phase, thanks to its Goldstone mode, has a higher bulk viscosity at
very low temperatures, we see that in the range
$10\,{\rm keV}\lesssim T \lesssim 10\,{\rm MeV}$ the CFL phase has a larger kaon-decay bulk viscosity than the CFL-$K^0$ phase.
This is because of the phase space available at the $K^0\leftrightarrow H$ ``resonance'' (see the end of Sec.~\ref{rates}).
There is another contribution to the bulk viscosity, from $H\leftrightarrow H+H$, which starts to
become comparable to the contribution from kaon decay at $T\lesssim 1\, {\rm MeV}$ \cite{Manuel:2007pz}.

The bulk viscosity of CFL/CFL-$K^0$ matter
is only comparable to that of unpaired quark matter at relatively high temperatures,
of order $10\,{\rm MeV}$. At lower temperatures $T\lesssim 5\,{\rm MeV}$ the bulk viscosity 
of CFL/CFL-$K^0$ quark matter is several orders of magnitude smaller. 
This is the case not only in comparison to the unpaired phase but 
to all other color-superconducting phases with ungapped fermionic modes---these
have bulk viscosities comparable to that 
of the unpaired phase, and even larger at high temperatures \cite{Alford:2006gy}.
The only other color-superconducting phase besides the CFL/CFL-$K^0$ phase in which all fermions may be gapped 
is the color-spin locked phase \cite{Schafer:2000tw,Schmitt:2004et,Marhauser:2006hy,Aguilera:2005tg}. For a discussion of its bulk viscosity 
see Ref.\ \cite{Sa'd:2006qv}.

\section{Summary and conclusions}
\label{conclusions}

We have computed the bulk viscosity of kaon-condensed CFL quark matter (CFL-$K^0$ phase). Kaon condensation affects the low-energy 
properties significantly and therefore has a significant effect on thermodynamic and transport properties of color-flavor locked
quark matter. In particular, the CFL-$K^0$ phase has a massless bosonic excitation associated with kaon condensation 
which is absent in the pure CFL phase. In both 
CFL-$K^0$ and CFL phases, there is also a massless Goldstone mode $H$ associated with superfluidity. For most of the thermodynamic
properties, the effect of the additional Goldstone mode is important, but rather easy to predict. We have shown this for the specific heat, 
which acquires a contribution of the kaon mode which, as expected, has the same temperature dependence as the contribution of the superfluid mode. 
The prefactor of the former, however, is typically larger than that of the latter such that the kaon 
contribution in fact dominates the specific heat at low temperatures. 

The effect of the additional Goldstone mode on the bulk viscosity is more complicated. We have used the results
of our earlier work \cite{Alford:2007qa} which provides a self-consistent description of the 
CFL-$K^0$ phase for arbitrary temperatures. Using the resulting thermal kaon mass and excitation energy, we have 
computed the density and the susceptibility of kaons, both of which are needed for the bulk viscosity. Moreover, we have
computed the rate of the processes $K^0\leftrightarrow H+H$ and $K^0+H\leftrightarrow H$, where we denote by $K^0$ both the neutral kaon 
in the CFL phase and the massless Goldstone mode arising upon kaon condensation in the CFL-$K^0$ phase. 
These weak processes serve to re-establish chemical (flavor) equilibrium
in response to an external volume 
oscillation, hence giving rise to a nonvanishing bulk viscosity. 

At very high temperatures, $T\gtrsim 10\,{\rm MeV}$, the difference in the kaon excitations in the CFL and CFL-$K^0$ phase 
is negligible. Consequently, in this case the kaon production (and annihilation) rate is almost identical for the two phases. 
At smaller (but not too small) temperatures, $10\,{\rm keV}\lesssim T\lesssim 10\,{\rm MeV}$, the masslessness
of the Goldstone mode in the CFL-$K^0$ phase suppresses this rate because of a smaller available phase space for the weak process. 
Since the timescale of the rates in both phases is     
smaller than the typical oscillation (and rotation) frequency in a compact star, this effect decreases the bulk viscosity of the CFL-$K^0$ phase
compared to the CFL phase. Another effect is given through the different susceptibilities. In the condensed system, the susceptibility 
at low temperatures is much larger than that of the uncondensed system. 
This effect works in the same direction, further decreasing the bulk viscosity compared to the uncondensed system. 
For even smaller temperatures, $T\lesssim 10\,{\rm keV}$ the phase space actually is {\em larger} in the CFL-$K^0$ phase and 
consequently the bulk viscosity is larger too. It is interesting to note that for the neutrino
emissivity the effect of kaon condensation is quite different: neutrino emissivity in the CFL-$K^0$ phase is larger than in the CFL phase for
all temperatures $T\lesssim 10\,{\rm MeV}$ \cite{Reddy:2003ap}.

We now have a fairly complete understanding of bulk viscosity
in color-flavor-locked phases of quark matter. The
suppression of the bulk viscosity due to the absence of ungapped fermionic excitations was
predicted in Ref.\ \cite{Madsen:1999ci}. Subsequent
more careful calculations took into account the contribution of the superfluid \cite{Manuel:2007pz}
and kaonic \cite{Alford:2007rw} Goldstone modes.
With the result of the present paper we have shown that the conclusion already drawn in Ref.\ \cite{Madsen:1999ci}
is, for temperatures $T\lesssim 1\,{\rm MeV}$, not changed by the contribution of the Goldstone modes: 
color-flavor locked quark matter, even in the presence 
of kaon condensation, has a much lower bulk viscosity 
than all other known phases of dense quark matter and than nuclear matter. Only at large temperatures, and thus in very 
young neutron stars, can the contribution of the Goldstone modes render the bulk viscosity comparable to that of unpaired quark matter.

We finally mention that besides the bulk (and shear) viscosity, other properties of color-superconducting quark matter 
also deserve attention. Its equation of state may be used to put constraints on the mass-radius relation of hybrid stars with a quark matter 
core and a hadronic mantle. These calculations rely on simple models whose 
parameters are poorly known in the strong-coupling region of interest. While NJL model calculations, mainly due to their relatively large 
predicted strange quark mass, tend to find no stable hybrid star with a CFL core \cite{Baldo:2002ju,Klahn:2006iw}, other parametrizations
of the equation of state allow for hybrid stars with masses compatible with the observations \cite{Alford:2004pf}.
Other observables that may distinguish between certain phases of 
color-superconducting quark matter or between quark matter 
and nuclear matter are for instance the cooling curve of the star or glitches (sudden spin-ups). The corresponding  
transport properties of color superconductors have already been computed in the literature, see for 
instance Refs.\ \cite{Jaikumar:2002vg,Schmitt:2005wg,Jaikumar:2005hy,Anglani:2006br} and 
Ref.\ \cite{Mannarelli:2007bs} for neutrino emissivity and shear modulus, respectively. It is an interesting and promising 
task for the future to extend these calculations and compare them with more and better astrophysical data in order to 
understand matter inside a compact star, and, ultimately, map out the phase diagram of cold and dense quark matter.  


\begin{acknowledgments}
The authors acknowledge valuable discussions with M.\ Huang, C.\ Manuel, H.\ Meyer, K.\ Rajagopal, 
A.\ Rebhan, I.\ Shovkovy, A.\ Steiner, and M.\ Tachibana, and support by the U.S. Department of Energy under contracts 
\#DE-FG02-91ER40628,  
\#DE-FG02-05ER41375 (OJI). 
A.S.\ ackowledges support by the FWF project P19526. 
\end{acknowledgments}

\appendix

\section{Specific heat in the 2PI formalism}
\label{AppA}

Here we derive the expression for the specific heat in the 2PI formalism. We have to take into account that the 
condensate $\phi$ as well as the full propagator $S$ are implicit functions of the temperature $T$ (remember
that $S$ contains the self-consistent thermal mass $M(T)$). With the definition (\ref{defcV}) we thus have to compute 
\be
c_V = T{\cal D}^2P \, , \qquad {\cal D}\equiv \frac{\partial S}{\partial T}\frac{\partial}{\partial S} +
\frac{\partial \phi}{\partial T}\frac{\partial }{\partial \phi} + 
\frac{\partial }{\partial T} \, ,
\ee
under the contraint that at the stationary point 
\begin{subequations} \label{constraint}
\bea
f_S&\equiv&\frac{\partial P}{\partial S}=0\, , \label{s1} \\
f_\phi &\equiv&\frac{\partial P}{\partial \phi}= 0 \, . \label{s2}
\eea
\end{subequations}
We find 
\be \label{D2p}
{\cal D}^2 P = \frac{\partial S}{\partial T}\left( \frac{\partial S}{\partial T}
\frac{\partial f_S}{\partial S} + \frac{\partial \phi}{\partial T}\frac{\partial f_S}{\partial \phi}
+\frac{\partial f_S}{\partial T}\right) + 
\frac{\partial \phi}{\partial T}\left( \frac{\partial S}{\partial T}
\frac{\partial f_\phi}{\partial S} + \frac{\partial \phi}{\partial T}\frac{\partial f_\phi}{\partial \phi}
+\frac{\partial f_\phi}{\partial T}\right) + 
\frac{\partial S}{\partial T}\frac{\partial f_S}{\partial T} + \frac{\partial \phi}{\partial T}
\frac{\partial f_\phi}{\partial T}+\frac{\partial^2 P}{\partial T^2} \, .
\ee
From the constraint (\ref{constraint}) we obtain 
\be
\left(\frac{\partial S}{\partial T},\frac{\partial \phi}{\partial T}\right) = -
\left(\frac{\partial f_S}{\partial T},\frac{\partial f_\phi}{\partial T}\right)
\left(\begin{array}{cc} \displaystyle{\frac{\partial f_S}{\partial S}} & \displaystyle{\frac{\partial f_\phi}{\partial S}} \\[2ex]
\displaystyle{\frac{\partial f_S}{\partial \phi}} &  \displaystyle{\frac{\partial f_\phi}{\partial \phi} }\end{array}\right)^{-1} \, . 
\ee
Inverting the matrix explicitly and inserting the result into Eq.\ (\ref{D2p}) shows that the expressions in parentheses vanish separately, 
and we are left with
\be 
c_V = T\,{\cal D} \frac{\partial P}{\partial T} \, .
\ee  
We thus have to apply the differential operator ${\cal D}$ on the explicit derivative of $P$ with respect to $T$.
From Eq.\ (\ref{pressure}) we find
\be \label{dpdt}
\frac{\partial P}{\partial T} = \int\frac{d^3{\bf k}}{(2\pi)^3}\left\{[1+f(\epsilon_k)]\ln[1+f(\epsilon_k)]-f(\epsilon_k)\ln f(\epsilon_k)\right\}
+\frac{1}{2}(M^2-m^2-2\alpha\phi^2-2\alpha I)\,\frac{\partial I}{\partial T} \, .
\ee
At the stationary point, the second term in this expression vanishes. This leaves the first term as the result for the entropy density.
In order to compute the second derivative, however, we have to keep the second term. 
By applying the differential operator ${\cal D}$ on this expression, the specific heat can be evaluated in a purely numerical way. 
For a further analytical evaluation we proceed as follows. 
In terms of the self-consistent quantities $\phi$ and $M$ we can write $c_V$ as (since $P$ only depends on the squares $\phi^2$ and 
$M^2$ we may take the derivatives with respect to the squares)
\be \label{cV1}
c_V = T\left(\frac{\partial^2 P}{\partial T\partial \phi^2} \frac{\partial \phi^2}{\partial T}
+\frac{\partial^2 P}{\partial T\partial M^2} \frac{\partial M^2}{\partial T}
+\frac{\partial^2 P}{\partial T^2}\right) \, .
\ee
The second derivatives of the pressure are straightforwardly obtained from Eq.\ (\ref{dpdt}). The derivatives of $\phi^2$ and $M^2$ 
with respect to the temperature depend on the stationarity equations. Below the critical temperature, one obtains the derivatives from 
Eqs.\ (\ref{stat1}) and (\ref{stat2}). Above the critical temperature, Eq.\ (\ref{stat1}) is automatically fulfilled
and the only equation is Eq.\ (\ref{stat2}) with $\phi$ set to zero. Consequently, the derivatives assume different functional 
forms below and above the critical temperature. We find
\begin{subequations} \label{dphiMdT}
\bea 
\frac{\partial \phi^2}{\partial T}&=&\left\{\begin{array}{cc} \displaystyle{-2\frac{\partial I}{\partial T}
\left(2\alpha\frac{\partial I}{\partial M^2}+1\right)^{-1}} & 
\mbox{for}\; T<T_c \, ,\\ & \\
0 & \mbox{for}\; T>T_c \, , \end{array}\right.  \\ [1ex]
\frac{\partial M^2}{\partial T}&=&\left\{\begin{array}{cc} \displaystyle{-2\alpha\frac{\partial I}{\partial T}
\left(2\alpha\frac{\partial I}{\partial M^2}+1\right)^{-1}} & 
\mbox{for}\; T<T_c \, ,\\ & \\
\displaystyle{-2\alpha\frac{\partial I}{\partial T}
\left(2\alpha\frac{\partial I}{\partial M^2}-1\right)^{-1}} & \mbox{for}\; T>T_c \, , \end{array}\right. \label{dMdT}
\eea
\end{subequations}
Now we can insert these expressions and the second derivatives of the pressure obtained from Eq.\ (\ref{dpdt}) into Eq.\ (\ref{cV1})
and evaluate the result at the stationary point, i.e., we use Eqs.\ (\ref{stationarity}). The result is
\be
c_V = \frac{1}{2}\int\frac{d^3{\bf k}}{(2\pi)^3}\frac{\epsilon_k^2}{T^2}\frac{1}{\cosh\frac{\epsilon_k}{T}-1}
-\frac{T}{2}\frac{\partial I}{\partial T}\frac{\partial M^2}{\partial T} \, .
\ee
With Eq.\ (\ref{dMdT}) one then obtains the results below and above the critical temperature. The discontinuity at $T=T_c$ then 
is
\be
\Delta c_V\equiv c_V(T\to T_c^+)-c_V(T\to T_c^-)=2\alpha T\left(\frac{\partial I}{\partial T}\right)^2
\frac{1}{1-4\alpha^2\left(\frac{\partial I}{\partial M^2}\right)^2} \, .
\ee

\section{Bulk viscosity with quark number effects}
\label{AppB}

In this appendix we derive the bulk viscosity taking into account the oscillations in the quark chemical potential as given in 
Eq.\ (\ref{mut2}). Instead of the simplified single differential equation (\ref{diff}) this yields the two coupled equations
\begin{subequations}
\bea
\frac{\partial n_q}{\partial\mu_q}\frac{\partial \mu_q}{\partial t}+\frac{\partial n_q}{\partial\mu}\frac{\partial \mu}{\partial t}&=&
\frac{\partial n_q}{\partial V}\frac{\partial V}{\partial t} \, , \\
\frac{\partial n}{\partial\mu_q}\frac{\partial \mu_q}{\partial t}+\frac{\partial n}{\partial\mu}
\frac{\partial \mu}{\partial t}&=&
\frac{\partial n}{\partial V}\frac{\partial V}{\partial t}-\lambda\,{\rm Re}(\delta\mu e^{i\omega t}) \, .
\eea
\end{subequations}
We insert the form of the volume oscillation (\ref{volume}) and Eqs.\ (\ref{dmut}) into these differential equations and find
the solution
\begin{subequations} \label{solution}
\bea
{\rm Im}\,\delta\mu_q &=& \frac{\delta V_0}{V_0}\omega\frac{\lambda(\frac{\partial n_q}{\partial\mu_q}n-\frac{\partial n}{\partial\mu_q}n_q)
}{({\rm det}J)^2\omega^2+(\frac{\partial n_q}{\partial\mu_q}\lambda)^2}\,\frac{\partial n_q}{\partial\mu}
\, , \\ 
{\rm Im}\,\delta\mu &=& -\frac{\delta V_0}{V_0}\omega\frac{\lambda(\frac{\partial n_q}{\partial\mu_q}n-\frac{\partial n}{\partial\mu_q}n_q)
}{({\rm det}J)^2\omega^2+(\frac{\partial n_q}{\partial\mu_q}\lambda)^2}\,\frac{\partial n_q}{\partial\mu_q}
\, ,
\eea
\end{subequations}
where $J$ denotes the Jacobian of the two-valued function $[n_q(\mu_q,\mu),n(\mu_q,\mu)]$,
\be \label{J}
J \equiv
 \left(\begin{array}{cc} \displaystyle{\frac{\partial n_q}{\partial\mu_q}} & \displaystyle{\frac{\partial n_q}{\partial\mu}}
\\[2ex] \displaystyle{\frac{\partial n}{\partial\mu_q}}& \displaystyle{\frac{\partial n}{\partial\mu}} \end{array}\right) \, .
\ee
Inserting Eqs.\ (\ref{solution}) into the definition of the bulk viscosity yields
\be \label{bulkgeneral}
\zeta = \left(n\frac{\partial \mu}{\partial n}+n_q\frac{\partial \mu_q}{\partial n}\right)
\left(n\frac{\partial \mu}{\partial n}+n_q\frac{\partial \mu}{\partial n_q}\right) \frac{\lambda}{\omega^2+
\left(\frac{\partial \mu}{\partial n}\lambda\right)^2} \, .
\ee
This is the general form of the bulk viscosity where the derivatives of the chemical potentials with respect to the densities are obtained
by inverting the Jacobian (\ref{J}). We now assume that the quark chemical potential enters the densities only through the kaon chemical 
potential, see Eq.\ (\ref{masschem}). This means that we neglect the dependence on $\mu_q$ through $f_\pi$ and $a$ in Eq.\ (\ref{matching}).
Defining the small dimensionless quantity
\be
\eta\equiv \frac{m_s^2}{2\mu_q^2} 
\ee
we then simply have $n_q=-n\,\eta$ and 
\be
J=\left(\begin{array}{cc} \displaystyle{\chi\eta^2+\frac{2n}{\mu_q}\eta} & -\chi\eta \\[2ex] -\chi\eta & \chi \end{array}\right) \, , \qquad 
J^{-1} = 
\left(\begin{array}{cc} \displaystyle{\frac{\partial \mu_q}{\partial n_q}} & \displaystyle{\frac{\partial \mu_q}{\partial n}} \\[2ex]
\displaystyle{\frac{\partial \mu}{\partial n_q}} & \displaystyle{\frac{\partial \mu}{\partial n}}\end{array}\right)
=\left(\begin{array}{cc} \displaystyle{\frac{\mu_q}{2n\eta}} &\displaystyle{\frac{\mu_q}{2n}} \\[2ex] \displaystyle{\frac{\mu_q}{2n}} & \displaystyle{\chi^{-1}+\frac{\mu_q}{2n}\eta} 
\end{array}\right) \, .
\ee
Consequently, Eq.\ (\ref{bulkgeneral}) becomes
\be \label{bulkfull}
\zeta = \frac{n^2}{\chi^2}\frac{\lambda}{\omega^2+
\left[\left(\chi^{-1}+\frac{\mu_q}{2n}\eta\right)\lambda\right]^2} \, .
\ee
Neglecting the term proportional to $\eta$ yields the result (\ref{bulkfinal}) that we use in the main part of the paper.

\section{Number susceptibility in the 2PI formalism}
\label{AppC}

In this appendix we derive an expression for the kaon number susceptibility $\chi$. Since $\chi$ is given by the second derivative 
of the pressure (\ref{pressure}) with respect to the chemical potential, we can use the same formalism as in Appendix \ref{AppA}, where 
we have computed the second derivative with respect to the temperature. Consequently, with the same arguments
as in Appendix \ref{AppA} (cf.\ Eq.\ (\ref{cV1}))
\be \label{chi1}
\chi = \frac{\partial^2 P}{\partial \mu\partial \phi^2} \frac{\partial \phi^2}{\partial \mu}
+\frac{\partial^2 P}{\partial \mu\partial M^2} \frac{\partial M^2}{\partial \mu}
+\frac{\partial^2 P}{\partial \mu^2} \, .
\ee
We determine the derivatives of $\phi^2$ and $M^2$ with respect to $\mu$ from the stationarity equations (\ref{stationarity}). They are
\begin{subequations} \label{dphiMdmu}
\bea 
\frac{\partial \phi^2}{\partial \mu}&=&\left\{\begin{array}{cc} \displaystyle{\frac{2\mu}{\alpha}-\frac{\phi^2}{\alpha}
\partial_\mu \alpha-\frac{2}{\alpha}\frac{\alpha\left(\partial_\mu I+
4\mu\partial_M I\right)+I\partial_\mu\alpha}{1+2\alpha\partial_M I}} & 
\quad\mbox{for}\; T<T_c \, ,\\ & \\
0 & \quad\mbox{for}\; T>T_c \, , \end{array}\right.  \\ [1ex]
\frac{\partial M^2}{\partial \mu}&=&\left\{\begin{array}{cc} \displaystyle{4\mu-2\frac{\alpha\left(\partial_\mu I+
4\mu\partial_M I\right)+I\partial_\mu\alpha}{1+2\alpha\partial_M I}}
& \quad\mbox{for}\; T<T_c \, ,\\ & \\
\displaystyle{2\frac{\alpha\partial_\mu I+
I\partial_\mu\alpha}
{1-2\alpha\partial_M I}}
& \quad\mbox{for}\; T>T_c \, , \end{array}\right. \label{dMdmu}
\eea
\end{subequations}
where we have abbreviated $\partial_\mu \equiv \partial /\partial\mu$, $\partial_M\equiv \partial/\partial M^2$.
Note that the effective coupling constant $\alpha$ depends on the kaon chemical potential $\mu$, see Eq.\ (\ref{alpha}). 
Next, we compute the second derivatives of the pressure appearing in Eq.\ (\ref{chi1}) with the help of Eq.\ (\ref{pressure}).
Putting everything together we obtain the susceptibility, 
\begin{subequations}
\bea
\chi&=& \left\{\begin{array}{cc} \displaystyle{\chi_0 +\chi_1 + \frac{1}{2T} \int\frac{d^3{\bf k}}{(2\pi)^3} \frac{1}{\cosh\frac{\epsilon_k}{T}-1}} & 
\quad\mbox{for}\; T<T_c \, ,\\ & \\
\displaystyle{\tilde{\chi}_1 + \frac{1}{2T} \int\frac{d^3{\bf k}}{(2\pi)^3} \frac{1}{\cosh\frac{\epsilon_k}{T}-1}} & \quad\mbox{for}\; T>T_c \, , \end{array}\right.
\eea
\end{subequations}
where the zero-temperature part is 
\bea
\chi_0 &\equiv& \phi^2+\frac{2\mu^2}{\alpha}\left(1-\frac{\phi^2}{2\mu}\partial_\mu \alpha
\right)^2 -\frac{\phi^4}{4}\partial^2_\mu \alpha \, , \label{chi0}  
\eea
and where 
\begin{subequations} \label{chi01}
\bea
\chi_1&\equiv& -\frac{(\partial_\mu I+4\mu\partial_M I)(2\mu-\alpha \partial_\mu I)}{1+2\alpha \partial_M I} - 2\mu\partial_\mu I -
I\partial_\mu \alpha\left(\frac{2\mu}{\alpha}-\frac{3\partial_\mu I+8\mu\partial_M I-2\mu/\alpha}{1+2\alpha \partial_M I}
\right) \non
&&+ \, (\partial_\mu \alpha)^2\frac{2I}{\alpha}\left(\phi^2+\frac{I}{1+2\alpha \partial_M I}
\right)-\phi^2 I\partial^2_\mu \alpha \, ,  \\
\tilde{\chi}_1&\equiv& -
\frac{\alpha \partial_\mu I+I\partial_\mu \alpha}{1-2\alpha \partial_M I}\partial_\mu I \, .
\eea
\end{subequations}

\section{Low-temperature result for the kaon rate}
\label{AppD}

In this appendix we derive Eq.\ (\ref{lambdasmallT}), the low-temperature expression for the kaon rate in the presence of a kaon condensate.
To this end, we first compute the low-temperature result for the imaginary part of the retarded $H$ self-energy. Below the critical 
temperature we have $\epsilon_p<vp$ for all $p$ (see Eq.\ (\ref{excite})) and thus we have 
\be
{\rm Im}\, \Pi(\epsilon_p,{\bf p}) = \Pi^-(\epsilon_p,{\bf p}) = \frac{\pi}{162\mu_q^4 v^7}\frac{\epsilon_p^2}{pf(\epsilon_p)}
\int_{\frac{vp+\epsilon_p}{2v}}^\infty dk\,g^2(\epsilon_p,p,k)f(vk)[1+f(vk-\epsilon_p)] \, .
\ee
For small temperatures only small momenta $p\lesssim T$ are relevant, and we can use the small-momentum approximation of the excitation energy
given in Eq.\ (\ref{lowk}). Moreover, since we consider temperatures well below the critical temperature, the 
quadratic part of the kaon dispersion is negligible. Therefore, $\epsilon_p\simeq \beta vp$, with the abbreviation 
\be \label{beta}
\beta\equiv\beta(m,\mu) \equiv \sqrt{\frac{\mu^2-m^2}{3\mu^2-m^2}} \, .
\ee
Here we have used the zero-temperature result for the self-consistent mass $M$. 
With the new integration variable $x=vk/T$ we have approximately
\be
{\rm Im}\, \Pi(\epsilon_p,{\bf p}) \simeq \frac{\pi\beta^2pT}{162\mu_q^4 v^6f(\epsilon_p)}
\int_{\frac{(1+\beta)vp}{2T}}^\infty dx\,[(\beta^2-1)v^2p^2-3(1-v^2)Tx(\beta vp-Tx)]^2\,e^{-x} \, .
\ee
Note that $p$ and $T$ are of the same order. Therefore we cannot simply drop the terms in the square brackets that are of higher
order in $T$. We rather have to keep all the terms. The integration can be done exactly and we obtain
\be
{\rm Im}\, \Pi(\epsilon_p,{\bf p}) \simeq \frac{\pi\beta^2pT}{216\mu_q^4 f(\epsilon_p)}\,e^{-\frac{(1+\beta)vp}{2T}}
[(\beta^2-1)^2p^4+8\sqrt{3}\,(\beta^2-1)\,p^3T+48\,(\beta^2+1)\,p^2T^2+576\sqrt{3}\,p\,T^3+3456\,T^4] \, , 
\ee
where, for the sake of brevity, we have used $v=1/\sqrt{3}$. We can now insert the result into the rate (\ref{Gammaapprox}). In the denominator 
of the integrand we neglect the term ${\rm Im}\,\Pi^2$. No further approximation is then required to obtain the final result. We make use 
of 
\be
\int_0^\infty dx\,\frac{x^ne^{-x}}{\sinh \beta x} = \frac{\Gamma(n+1)}{2^n\beta^{n+1}}\,\zeta\left(n+1,\frac{1+\beta}{2\beta}\right) \, , 
\ee
where $\zeta$ is the generalized zeta function, and where we need the cases $n=2,\ldots, 6$. Then we arrive at 
\be \label{lQ}
\lambda \simeq \frac{9G_{ds}^2f_\pi^2f_H^2}{2\pi}\,\frac{\beta(\beta^2-v^2)^2}{(\beta^2-1)^2}Q(\beta)\frac{T^7}{\mu_q^4} \, , 
\ee
with the dimensionless function
\bea \label{Q}
Q[\beta(m,\mu)] &\equiv& 15\,\frac{(\beta^2-1)^2}{\beta^7}\,\zeta\left(7,\frac{1+\beta}{2\beta}\right)
+20\,\frac{\beta^2-1}{\beta^6}\,\zeta\left(6,\frac{1+\beta}{2\beta}\right)+8\,\frac{\beta^2+1}{\beta^5}\,
\zeta\left(5,\frac{1+\beta}{2\beta}\right) \non
&& +\;
24\,\frac{1}{\beta^4}\,\zeta\left(4,\frac{1+\beta}{2\beta}\right)+16\,\frac{1}{\beta^3}\,\zeta\left(3,\frac{1+\beta}{2\beta}\right) \, , 
\eea
Inserting the expression for $\beta$ from Eq.\ (\ref{beta}) into Eq.\ (\ref{lQ}) yields the final result 
given in the main text, see Eq.\ (\ref{lambdasmallT}).

\bibliography{refs}

\end{document}